\documentclass[12pt]{article}
\usepackage{newtxtext,newtxmath}
\usepackage{graphicx}
\usepackage[letterpaper,margin=1in]{geometry}
\renewenvironment{abstract}
	{\quotation}
	{\endquotation}

\date{}

\makeatletter
\renewcommand{\fnum@figure}{\textbf{Figure \thefigure}}
\renewcommand{\fnum@table}{\textbf{Table \thetable}}
\makeatother
\usepackage{scicite}
\usepackage{url}
\def\scititle{
	Dynamical universality in a driven quantum fluid of light
}
\title{\bfseries \boldmath \scititle}
\author{
	I.~Gnusov$^{1\dagger}$,
	P.~Comaron$^{1\ast\dagger}$,
    A.~Gianfrate$^{1}$,
    D.~Trypogeorgos$^{1}$,
    M.~H.~Szymanska$^{2}$,\\
    P.~Cazzato$^{1}$,
    M.~De Giorgi$^{1}$,
    D.~Sanvitto$^{1\ast}$,
    D.~Ballarini$^{1}$\and
	\small$^{1}$CNR Nanotec, Institute of Nanotechnology, via Monteroni, 73100, Lecce, Italy\and
	\small$^{2}$Department of Physics and Astronomy, University College London, \\ \small Gower Street, London, WC1E 6BT, United Kingdom.\and
	\small$^\ast$Corresponding author. Email: paolo.comaron@cnr.it\and
    \small$^\ast$Corresponding author. Email: daniele.sanvitto@cnr.it\and
	\small$^\dagger$These authors contributed equally to this work.
}

\begin{document} 

\maketitle
\begin{abstract} \bfseries \boldmath
Universal scaling near phase transitions is one of the central ideas of physics, linking the growth of spatial correlations to the slowing down of dynamics. So far, direct experimental access to this critical behavior has remained largely confined to equilibrium many-body systems, and especially to static critical behavior. Here we probe how universality emerges in a driven quantum fluid of light formed by exciton--polaritons in a semiconductor microcavity. By probing the fluctuation-dominated disordered phase below the condensation threshold, we directly measure both the static growth of the correlation length $\xi$ and the dynamical slowing down of the relaxation time $\tau$. We find that these quantities obey the universal relation $\tau \propto \xi^{z}$ with dynamical exponent $z \approx 2$, revealing diffusive dynamics of a non-conserved order parameter. Our results extend the physics of critical dynamics from equilibrium matter to driven optical systems, bridging quantum condensates and lasers.
\end{abstract}

\noindent

\

Critical phenomena arise from the growth and slow relaxation of fluctuations as a many-body system approaches a phase transition. Near the critical point, long-wavelength fluctuations dominate the dynamics, leading to the divergence of the correlation length and to the phenomenon of critical slowing down.
These behaviors are largely independent of microscopic details and fall into a small number of dynamical universality classes introduced by Halperin and Hohenberg~\cite{hohenberg1977theory}. Within each class, disparate physical systems exhibit the same scaling of spatial and temporal correlations
\begin{align}
\xi(\varepsilon) &\propto |\varepsilon|^{-\nu}, \label{eq:xi_epsilon_} \\
\tau(\varepsilon) &\propto \xi^{\,z} \propto |\varepsilon|^{-\nu z}, \label{eq:tau_epsilon}
\end{align}
where the reduced control parameter $\varepsilon = 1 - \lambda/\lambda_c$ measures the distance of the control parameter $\lambda$ from the critical point $\lambda_c$, while $\nu$ and $z$ are the static and dynamic critical exponents.
At equilibrium, detailed balance constrains the relaxation of fluctuations, providing the framework in which dynamical universality classes and their associated exponent $z$ are defined. Whether an analogous level of universality persists in systems operating out of equilibrium remains an open question. Driven--dissipative systems are continuously coupled to external reservoirs, so that steady states are selected by a balance between gain and loss rather than by thermal equilibration, and the connection to equilibrium statistical mechanics becomes nontrivial~\cite{Tauber_2014, Sieberer2025RMP, TauberPRX2014}.

Optical systems provide a particularly appealing platform to explore these questions. The laser threshold is one of the earliest examples of a non-equilibrium phase transition, associated with the spontaneous emergence of a macroscopic coherent field~\cite{graham-haken-1970, degiorgio-scully-1970, 1970quop.conf..201H, haken-laser-1975}. Yet, despite decades of development, spatially extended lasers have rarely been used to probe the universal properties of critical fluctuations. Most studies have instead focused on single-mode cavities or on steady-state coherence above threshold, leaving largely unexplored the regime where critical slowing down and growing correlations occur.

Quantum fluids of light realized with exciton--polaritons in semiconductor microcavities provide a bridge between atomic condensates and laser physics~\cite{carusotto2013quantum, szymanska2006nonequilibrium}. Like atomic Bose--Einstein condensates, polariton systems form spatially extended coherent states and exhibit collective phenomena such as quantum vortices~\cite{Lagoudakis2008firtsv, Dominici2015} and superfluid behavior~\cite{lerario_room-temperature_2017}, while operating in a non-equilibrium steady state sustained by continuous pumping and dissipation. Above threshold, polariton condensates have revealed Berezinskii--Kosterlitz--Thouless physics in two dimensions (2D)~\cite{caputo2018topological} and Kardar--Parisi--Zhang scaling in 1D and 2D driven phase dynamics~\cite{Fontaine2022,widmann2026_2D_KPZ}. However, a direct experimental determination of both static and dynamical critical scaling remains lacking. Such scaling must be probed below threshold, where the order parameter vanishes and the system dynamics is governed by the growth and relaxation of fluctuations. Accessing this regime is particularly challenging because temporal correlations are not directly encoded in steady-state observables and, in driven--dissipative systems, can be obscured by additional timescales associated with gain, losses, and coupling to external reservoirs.

Here we experimentally investigate critical fluctuations in a driven quantum fluid of light formed by exciton--polaritons in semiconductor microcavities. Combining interferometric measurements with controlled perturbations of the condensate, we directly access the scaling relation linking spatial and temporal correlations. This approach allows us to determine the dynamical scaling of the system $\tau \propto \xi^{z}$ near criticality. 
Our results show that dynamical universality can emerge in a driven–dissipative quantum system despite the absence of detailed balance, and connect it to the long-standing interpretation of the laser threshold as a non-equilibrium phase transition.

\subsection*{2D exciton polariton condensates}

Above threshold, polariton condensates develop a finite order parameter and spontaneously break the $U(1)$ symmetry. The resulting Goldstone mode corresponds to long-wavelength phase fluctuations, which in driven--dissipative systems are diffusive or weakly propagating~\cite{szymanska2007mean, wouters2007, claude2025goldstone, ballarini2020directional, stepanov2019dispersion}. In this regime, the dynamics is dominated by phase modes of an already established coherent field.

Below threshold, by contrast, the system remains in the symmetric phase with vanishing order parameter in the thermodynamic limit. Here the relevant degrees of freedom are fluctuations of the full complex field $\psi$, involving both amplitude and phase. Approaching the critical point, the effective relaxation rate of these fluctuations vanishes and long-wavelength modes become critically slow.

Polariton condensates under non-resonant pumping are commonly described by a driven--dissipative Gross--Pitaevskii or stochastic complex Ginzburg--Landau equation coupled to an excitonic reservoir (see Supplementary Materials for details)~\cite{chiocchetta2013,woutersLiew2010,wouters2010,comaron2018dynamical,zamora2020}. 
Below threshold, where the order parameter remains small, the reservoir can be adiabatically eliminated and the dynamics linearized around the symmetric state $\langle \psi \rangle = 0$, where $\psi$ describes the critical fluctuations of the order parameter, yielding an effective equation:
\begin{equation}
\partial_t \psi =
r \psi + \mathcal{D}\nabla^2\psi +  i \mathcal{L} \psi  + \eta,
\label{eq:diffusive}
\end{equation}
where $r = \gamma_\mathrm{c} \varepsilon / 2$ is the control parameter set by the reduced pump $\varepsilon = \mathcal{P}/\mathcal{P}_\mathrm{th}-1$, which measures the distance from the critical point, with $\mathcal{P}$ the pump strength and $\mathcal{P}_\mathrm{th}$ its value at criticality, and $\gamma_\mathrm{c}$ is the inverse of the polariton lifetime. $\mathcal{D}$ is an effective diffusion coefficient~\cite{wouters2012energyrelax}, the term $i\mathcal{L}\psi$ generates coherent phase rotation and dispersion, and $\eta$ is a noise term originating from gain and loss processes. 
At long wavelengths, this equation suggests diffusive relaxational critical dynamics of a non-conserved complex field. The central question is whether such a scaling regime remains observable in the full physical system, despite its driven--dissipative character and finite size~\cite{TauberPRX2014, Sieberer2025RMP}. 
In the following, we address this question by measuring independently the spatial and temporal correlations of fluctuations below threshold and by comparing the results with simulations of the full driven--dissipative model.

\subsection*{Spatial and temporal slowing down}

\begin{figure}
\centering
\includegraphics[width=.8\linewidth]{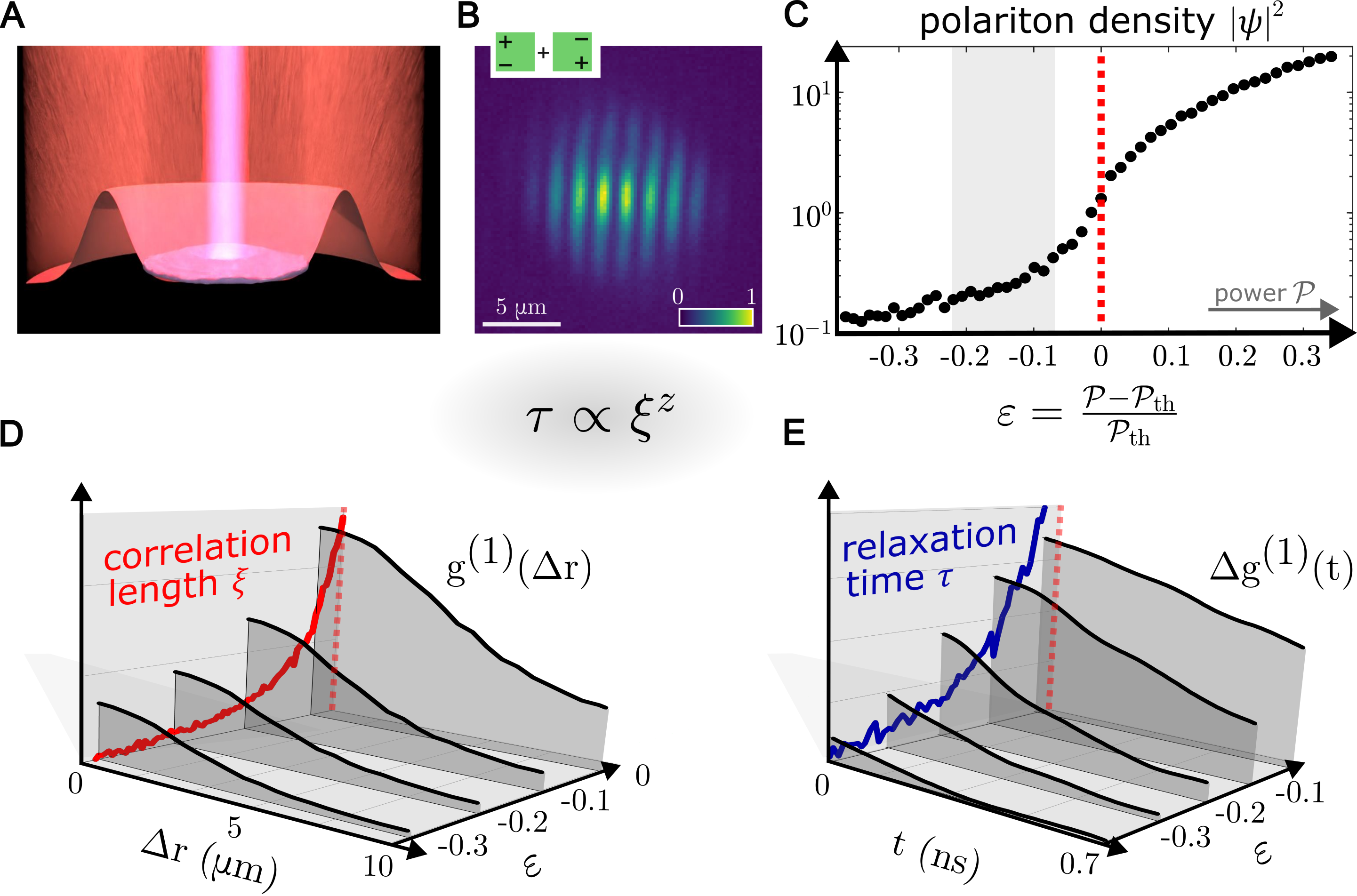}
\caption{\textbf{Experimental configuration}. \textbf{A} Schematic of polariton condensate formed in an optical trap. \textbf{B} Interference fringes of the polariton field interfered with retro-reflected copy of itself ($\varepsilon = -0.05$). \textbf{C} Polariton density as a function of reduced pump $\varepsilon$, with the red dashed line indicating $\varepsilon=0$.  
\textbf{D} Critical divergence of the correlation length $\xi$ (red curve) as a function of $\varepsilon$. Black curves represent the correlation function $g^{(1)}(\Delta r)$ as measured experimentally. 
\textbf{E} Critical slowing down of the relaxation time $\tau$ (blue curve) as a function of reduced power ($\varepsilon$). Black curves depict the decay $\Delta g^{(1)}(t)$ of the weak perturbation. The dynamical exponent is then extracted from the relation $\tau \propto \xi^z$ for $\varepsilon \to 0$.}
\label{fig:fig1}
\end{figure}

To probe static and dynamical critical behavior, we measure spatial and temporal coherence of a polariton condensate confined in a ring trap under non-resonant continuous-wave pumping $\mathcal{P}$, as sketched in Fig.~\ref{fig:fig1}\textbf{A}. Under non-resonant pumping, the condensate is fed by an excitonic reservoir whose relaxation dynamics could, in principle, introduce an additional timescale and affect the measured critical slowing down. This geometry offers two key advantages: it reduces the condensate–reservoir coupling and suppresses multimode condensation, favoring occupation of the lowest-energy state~\cite{Askitopoulos2013, Alnatah2024}. Using optical interferometry, we directly access the spatial correlator, as shown in Fig.~\ref{fig:fig1}\textbf{B} and detailed in Supplementary Materials.

Figure~\ref{fig:fig1}\textbf{C} shows the total polariton density as a function of pump power $\mathcal{P}$. The critical point, marked by the red line in Fig.~\ref{fig:fig1}\textbf{C}, is identified as the power at which the second derivative of the density changes sign on a log--log scale, as further discussed in Supplementary Materials. 

We then extract the correlation length $\xi(\varepsilon)$, obtained by fitting the spatial correlator as $g^{(1)}(\bf{r}) \propto e^{-|\bf{r}|/\xi(\varepsilon)}$, for a range of values $\varepsilon<0$ (dashed area in Fig.~\ref{fig:fig1}\textbf{C}). As the threshold is approached from below, the correlation length increases strongly, as illustrated in Fig.~1\textbf{D}. 
To probe the temporal dynamics, we perturb the steady state with a weak pulse and extract the relaxation time $\tau$ from the exponential decay of the spatially integrated correlator, $\propto e^{-t/\tau}$. The relaxation time also increases as the threshold is approached, as shown in Fig.~1\textbf{E}. The dynamical exponent $z$ is then obtained by comparing the spatial and temporal scales through the scaling relation $\tau \propto \xi^{z}$.

\begin{figure}
\includegraphics[width=\linewidth]{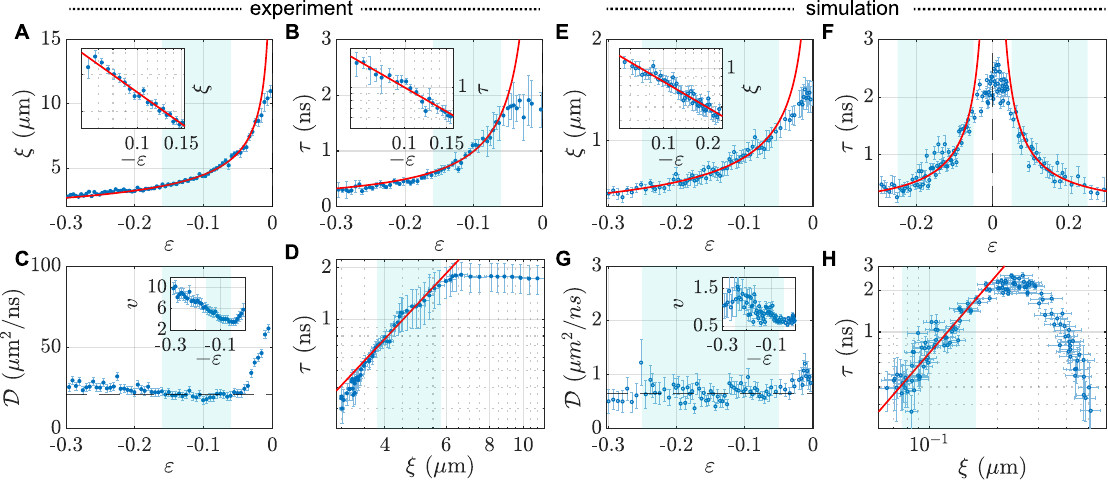}
\caption{\textbf{Spatial and temporal critical slowing down}.  
\textbf{A} Experimental and \textbf{E} numerical coherence length $\xi$ as a function of the reduced pump $\varepsilon$, approaching threshold from below. \textbf{B} Experimental and \textbf{F} numerical relaxation time $\tau$ of the perturbation as a function of $\varepsilon$. Insets in panels \textbf{A}, \textbf{B}, and \textbf{E} show the corresponding data on log--log scales. \textbf{C} Experimental and \textbf{G} numerical diffusion coefficient $\mathcal{D}$ as a function of $\varepsilon$. Insets in panels \textbf{C} and \textbf{G} show the corresponding slowing-down rate $v$ toward the critical point. 
\textbf{D} Experimental and \textbf{H} numerical relaxation time $\tau$ as a function of $\xi$ on log--log scales. Red lines are power-law fits. 
The light-blue regions indicate the fitting windows. The optical trap diameter is $15.5~\mu\mathrm{m}$ for the experiments and  $13.6~\mu\mathrm{m}$ for the simulations.
We use $|X|^2 = 0.41$ for the experiments and $|X|^2 = 0.5$ for the simulations. 
}
\label{fig:fig2}
\end{figure}

Figures~2\textbf{A} and 2\textbf{B} show the measured dependence of $\xi$ and $\tau$ on the reduced pump parameter $\varepsilon$, fitted to the power laws $\xi \propto |\varepsilon|^{-\nu}$ and $\tau \propto |\varepsilon|^{-\nu z}$, respectively.
As the critical point is approached from the disordered phase, $\tau$ grows faster than $\xi$, consistent with critical slowing down.
From Eqs.~\eqref{eq:xi_epsilon_}, ~\eqref{eq:tau_epsilon} and ~\eqref{eq:diffusive}, one can define an effective diffusion coefficient $\mathcal{D}$ and a characteristic propagation rate of correlations $v$ as 
\begin{equation}
\mathcal{D} = \frac{\xi^2}{\tau} = |\varepsilon|^{-\nu(2-z)}, \qquad v \propto \frac{\xi}{\tau} \propto |\varepsilon|^{\nu(z-1)}.
\end{equation}
We plot $\mathcal{D}$ and $v$ as a function of $\varepsilon$ in Fig.~\ref{fig:fig2}\textbf{C}, and in its inset, respectively. 
The approximately constant value of $\mathcal{D}$ in the shaded region below threshold is consistent with diffusive relaxation dynamics.
The corresponding decrease of $v$ on approaching threshold provides an additional indication of the slowing down of the critical fluctuations. 

A more robust determination of the dynamical exponent $z$, independent of the precise location of the critical point, is obtained by directly probing the scaling relation between relaxation time and correlation length, $\tau \propto \xi^{z}$, as shown in Fig.~\ref{fig:fig2}\textbf{D}. 

Numerical simulations of the full stochastic driven--dissipative model are reported in Fig.~\ref{fig:fig2}\textbf{E}, \textbf{F},\textbf{G}, \textbf{H} and show good agreement with the experimental observations. 

\subsection*{Finite size effects, detuning, and resonant control}

\begin{figure}[t]
\centering
\includegraphics[width=.7\linewidth]{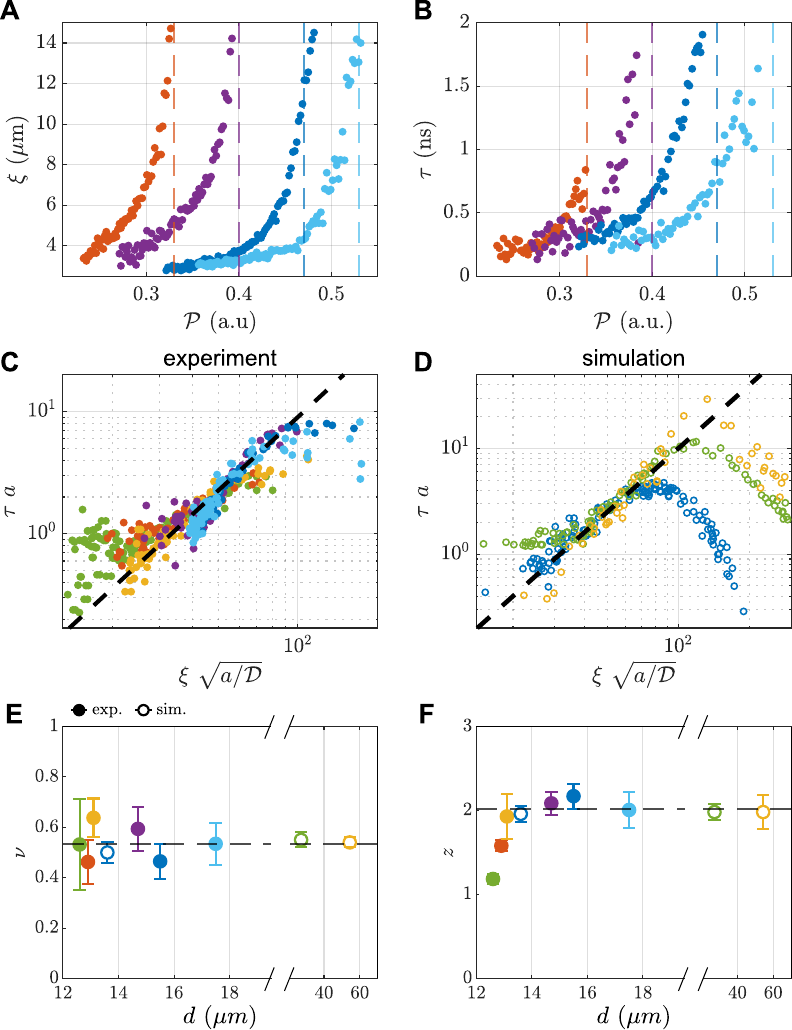}
\caption{\textbf{Size variation}.  Experimentally measured \textbf{A} coherence length $\xi$ and \textbf{B} relaxation time $\tau$ as a function of absolute pump power $P$ in arbitrary units for the optical trap with diameter d = $12.9 , 14.7, 15.5$ and $17.5 \mu m$ corresponding to the red, purple, dark blue and light blue points, respectively. Rescaled dependence of $\tau$ on $\xi$ extracted from \textbf{C} experiment and  \textbf{D} numerical simulations. The diffusion coefficient $\mathcal{D}$ is directly obtained from experiments and simulations, while $a = \gamma_c/2$ is the characteristic microscopic rate. \textbf{E} Static $\nu$ and  \textbf{F} dynamical $z$ critical exponents for different sizes of the optical trap extracted from the experiment (closed circles) and simulation (open circles). Dashed horizontal lines indicate the values $\nu = 0.54$ and $z = 2.02$, respectively, obtained averaging experiments and numerics, after excluding the two smallest experimental traps.
}
\label{fig:fig3}
\end{figure}

To test whether the observed scaling reflects an intrinsic property of the polariton field rather than a feature of a specific finite geometry, we vary the system size by tuning the diameter of the non-resonant ring excitation. 
Figure~\ref{fig:fig3}\textbf{A},\textbf{B} reports the correlation length $\xi$ and the relaxation time $\tau$ as a function of the pump power for different trap sizes. 
Increasing the trap diameter shifts the threshold to higher pump powers and modifies the non-universal coefficients $D$ and $v$. Nevertheless, all curves $\xi(\varepsilon)$ and $\tau(\varepsilon)$ display the same scaling behavior in the pre-asymptotic region approaching threshold. This shows that changing the finite-size cutoff affects the microscopic scales of the problem, but does not destroy the critical scaling regime. 

Upon rescaling $\xi$ by $\sqrt{\mathcal{D}/a}$ and $\tau$ by the characteristic microscopic timescale $a^{-1}$, all data collapse onto the same power-law relation $\tau \propto \xi^{z}$, confirming the universal scaling (Fig.~\ref{fig:fig3}\textbf{C}, \textbf{D}).
Here $a = \gamma_\mathrm{c}/2$ (see Supplemental Materials for derivation) represents a microscopic rate, setting the natural timescale of the problem and relating the reduced pump $\varepsilon$ to the effective growth rate $r$ (see Eq.~\eqref{eq:diffusive}).

The collapse in Fig.~\ref{fig:fig3} \textbf{C}, \textbf{D} therefore indicates that finite-size effects mainly renormalize the diffusion coefficient $\mathcal{D}$, while leaving the dynamical exponent $z$ unchanged. 

Full details of the numerical analysis and a quantitative estimate of the finite-size corrections are provided in the Supplementary Materials~\cite{campostrinivicari2009}.

In optical traps with diameter larger than 18 $\mu m $, condensation into first excited state occurs below the pump threshold for condensation into the lowest energy state, complicating the analysis of the coherence properties~\cite{Aladinskaia2023}. In numerical simulations, larger trap sizes remain accessible thanks to longer polariton lifetime, reaching diameters $L > 20\xi$ without noticeable changes in the extracted exponents.
Additional confirmation is obtained using periodic boundary conditions. 

The extracted critical exponents for the different trap sizes are summarized in Fig.~\ref{fig:fig3}(e,f), shown as filled and open symbols for experiments and simulations, respectively. Both yield values consistent with {$\nu = 0.56 \pm 0.08$ and $z = 2.04 \pm 0.19$ for the experiment, and $\nu = 0.53 \pm 0.02$ and $z = 1.97 \pm 0.08$ for the numerics}.

At the same time, the smallest traps show a tendency towards lower effective values of $z$, approaching $z\to 1$. This indicates the onset of crossover effects, which can arise either from finite-size constraints or from the increased influence of the excitonic reservoir. Indeed, in smaller traps, the spatial overlap between the condensate and the reservoir is enhanced, leading to stronger reservoir--polariton interactions and a consequent modification of coherence dynamics. 
This interpretation is supported by complementary measurements performed under Gaussian non-resonant excitation, where the condensate is formed directly on top of the reservoir and the overlap is maximal. 
In that configuration, we do not observe any significant growth of the relaxation time $\tau$ approaching threshold, indicating that the dynamics is then dominated by reservoir-controlled relaxation rather than by intrinsic critical slowing down of the polariton field  (see Supplementary Materials).

\begin{figure}[t!]
\centering
\includegraphics[width=\linewidth]{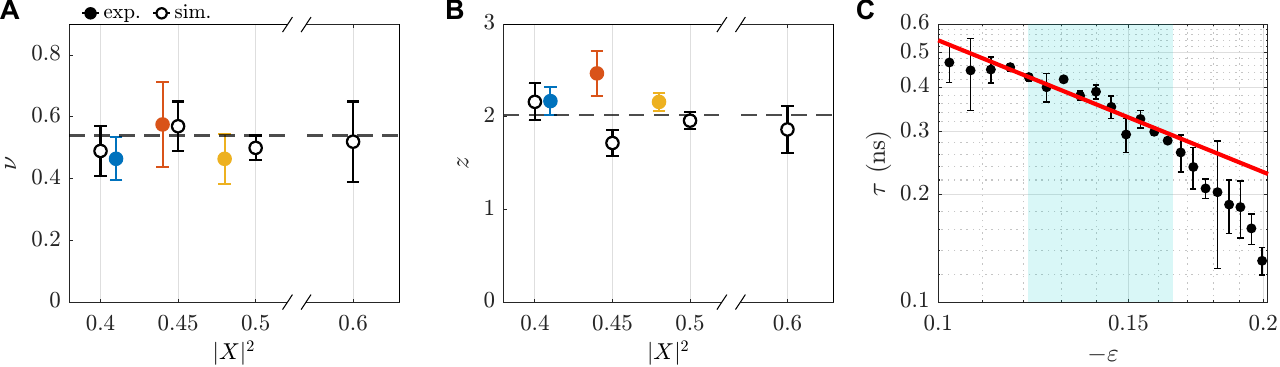}
\caption{\textbf{Detuning dependence and resonant perturbation.}
\textbf{A} Static exponent $\nu$ and \textbf{B} dynamical exponent $z$ as a function of the excitonic fraction $|X|^2$, extracted from experiments (closed circles) and numerical simulations (open circles). Dashed horizontal lines indicate the averaged values $\nu = 0.54$ and $z = 2.02$, respectively, as reported in Fig.~3. \textbf{C} Relaxation time $\tau$ as a function of the reduced pump $-\varepsilon$ on a log--log scale, measured with a resonant perturbation. The red line is a power-law fit yielding $z \simeq 2$. The optical trap diameter is $15.5~\mu\mathrm{m}$ for the experiments and $13.6~\mu\mathrm{m}$ for the simulations.
}
\label{fig:fig4}
\end{figure}

To further disentangle the observed scaling from microscopic parameters, we extract the critical exponents for different excitonic fractions. Changing the exciton fraction of polaritons modifies both the interaction strength and the coupling to the excitonic reservoir.
The universal diffusive scaling reported here is experimentally confirmed up to excitonic fraction $|X|^2 = 0.48$. For larger excitonic fractions, the present data do not allow the same scaling regime to be isolated as clearly.
Numerical simulations incorporating different interaction parameters ($g$, $g_\mathrm{R}$) and scattering rate $R$ reproduce the same trend and remain consistent with a diffusive scaling $z \simeq 2$, in good agreement with the experimental results shown in Fig.~\ref{fig:fig4} \textbf{A}, \textbf{B}. 

Finally, to test directly whether the slow decay measured in the time-resolved experiment belongs to the polariton field itself rather than to the reservoir, we perform complementary measurements using a coherent resonant probe. In this configuration, the perturbation directly seeds the polariton field without creating an additional non-resonant reservoir, providing a cleaner access to the intrinsic relaxation dynamics. Owing to the finite free spectral range of the pulsed laser, these measurements cannot be repeated for different trap sizes. Nevertheless, by tuning the probe close to the condensate energy below threshold and varying the continuous-wave pump power, we again observe a clear power-law dependence of the relaxation time (Fig.~\ref{fig:fig4}\textbf{C}), yielding a dynamical exponent consistent with $z \simeq 2$. This agreement with the non-resonant measurements provides strong evidence that the observed scaling is not set by the reservoir dynamics, but reflects the intrinsic critical relaxation of the polariton field.

\subsection*{Discussion and Conclusions}

Our results demonstrate dynamical scaling in a driven quantum fluid of light. By measuring both the correlation length and the relaxation time below threshold, we establish the scaling relation $\tau \propto \xi^{z}$ with $z \simeq 2$, identifying a diffusive critical regime in a non-equilibrium condensate. This provides direct experimental access to the dynamics of critical fluctuations in a regime that has remained largely unexplored experimentally. 
The collapse $\tau(\xi)$ persists across variations in trap size, excitonic fraction, and numerical realizations, showing that the observed dynamics is governed by a robust scaling structure rather than by microscopic details of a specific experimental configuration. At the same time, the agreement between resonant and non-resonant perturbations shows that the slow relaxation is intrinsic to the polariton field rather than being set by the excitonic reservoir.

More broadly, exciton--polariton condensates lie at the crossroads of Bose--Einstein condensation and laser physics~\cite{chiocchetta_gambassi_carusotto_2017}. In equilibrium quantum fluids with conserved particle number, critical dynamics is shaped by conservation laws and hydrodynamic coupling, and is naturally discussed within the Halperin--Hohenberg framework, typically in regimes of Model-F type~\cite{Donner2007Science,Navon2015science}. In driven systems such as polariton condensates or lasers, by contrast, no equally general classification is available. In this context, our results show that the driven polariton fluid nevertheless realizes a diffusive critical regime consistent with relaxational dynamics of a non-conserved complex order parameter. In this sense, polariton condensates provide an experimentally accessible setting in which Model-A--like dynamical scaling can emerge beyond the equilibrium framework. 

The measured static exponent {$\nu = 0.56 \pm 0.08$} is close to the Gaussian value $\nu = 1/2$ expected for relaxational dynamics of a non-conserved complex order parameter~\cite{goldenfeld2018lectures,delCampo2013,kleebank2025phot_crit_scaling}, while leaving open the possibility of deviations closer to the asymptotic critical point~\cite{TauberPRX2014}. 

Numerical simulations of the full stochastic driven--dissipative model support this interpretation and show that microscopic ingredients such as confinement, reservoir coupling, and energy relaxation primarily modify non-universal quantities, including $\xi$, $\tau$, and the effective diffusion coefficient $D$, without changing the scaling relation itself. Finite-size effects therefore renormalize the microscopic scales of the problem rather than the dynamical exponent.

At the same time, the universal regime identified here has clear boundaries. For very small traps, finite-size effects progressively obscure the clean diffusive scaling reported in the main text. 
Extending these measurements to larger systems and closer proximity to threshold will therefore be essential to probe the asymptotic regime beyond the pre-asymptotic window accessed here. Moreover, exploring a broader range of experimental parameters, including larger excitonic fractions, constitutes a natural direction for future work. 

Finally, our observations extend the long-standing interpretation of the laser threshold as a non-equilibrium phase transition to a regime where critical fluctuations can be resolved directly in both space and time. The key advance is therefore the demonstration that a finite, realistic, and genuinely driven quantum system can nevertheless exhibit a well-defined dynamical scaling regime. This establishes quantum fluids of light as a platform where universality can be tested experimentally beyond the equilibrium paradigm.

\newpage

\newpage

\section*{Acknowledgments}
We thank L. Canet, A. Minguzzi, H. Weinberger and M. Weitz for fruitful discussions.
\paragraph*{Funding:}
We acknowledge financial support from “Quantum Optical Networks based on Exciton-polaritons” (Q-ONE, N. 101115575, HORIZON - EIC - 2022 - PATHFINDER CHALLENGES EU project), "Neuromorphic Polariton Accelerator" (PolArt, N.101130304, Horizon-EIC-2023-Pathfinder Open EU project), “National Quantum Science and Technology Institute” (NQSTI, N. PE0000023, PNRR MUR project), “Integrated Infrastructure Initiative in Photonic and Quantum Sciences” (I-PHOQS, N. IR0000016, PNRR MUR project)
Views and opinions expressed are, however, those of the author(s) only and do not necessarily reflect those of the European Union or European Innovation Council and SMEs Executive Agency (EISMEA). Neither the European Union nor the granting authority can be held responsible for them.
M. H. S. acknowledges financial support from EPSRC (Grants No. EP/V026496/1).
This study was conducted using the DARIAH HPC cluster at CNR-NANOTEC in Lecce, funded by the ``MUR PON Ricerca e Innovazione 2014-2020" project, code PIR01$\_$00022 and H2IOSC Project  - Humanities and cultural Heritage Italian Open Science Cloud funded by the European Union – NextGenerationEU – NRRP M4C2 - Project code IR0000029. We thank G. Marra and S. Pascali for computational assistance.
\paragraph*{Author contributions:}
I.G. performed the experimental measurements, and P.Co. carried out the numerical simulations. I.G., P.Co and D.B. analysed the data. MHS provided theoretical support, A.G., D.T., M.D.G. and P.Ca. provided experimental support. P.Co., D.S., and D.B. conceived the work. All authors contributed to the interpretation of the results and to the discussion of the manuscript.
\paragraph*{Competing interests:}
There are no competing interests to declare.
\paragraph*{Data and materials availability:}
The data, software and simulation outputs will be made available through a public repository upon publication.

\newpage

\renewcommand{\thefigure}{S\arabic{figure}}
\renewcommand{\thetable}{S\arabic{table}}
\renewcommand{\theequation}{S\arabic{equation}}
\renewcommand{\thepage}{S\arabic{page}}
\setcounter{figure}{0}
\setcounter{table}{0}
\setcounter{equation}{0}
\setcounter{page}{1} 
\begin{center}
\section*{Supplementary Materials for:\\ \scititle}

	I.~Gnusov$^{1\dagger}$,
	P.~Comaron$^{1\ast\dagger}$,
    A.~Gianfrate$^{1}$,
    D.~Trypogeorgos$^{1}$,
    M.~H.~Szymanska$^{2}$,\\
    P.~Cazzato$^{1}$,
    M.~De Giorgi$^{1}$,
    D.~Sanvitto$^{1\ast}$,
    D.~Ballarini$^{1}$ \\
	
	\small$^{1}$CNR Nanotec, Institute of Nanotechnology, via Monteroni, 73100, Lecce, Italy\\
	\small$^{2}$Department of Physics and Astronomy, University College London, \\ \small Gower Street, London, WC1E 6BT, United Kingdom.\\	
	\small$^\ast$Corresponding author. Email: paolo.comaron@cnr.it\\
    \small$^\ast$Corresponding author. Email: daniele.sanvitto@cnr.it\\
	\small$^\dagger$These authors contributed equally to this work.

\end{center}

\subsubsection*{This PDF file includes:}
Materials and Methods\\
Supplementary Text\\
Figures S1 to S14\\
Tables S1 to S2\\

\newpage


\subsection*{Materials and Methods}

Experiments are performed on a planar $3/2\,\lambda$ $GaAs/AlGaAs$ microcavity containing 12 $GaAs$ quantum wells distributed at the antinodes of the confined electromagnetic field, with a collective Rabi coupling of 8 meV. The top distributed Bragg reflector (DBR) consists of 34 pairs of $AlAs/Al_{0.2}Ga_{0.8}As$ layers, while the back reflector contains 40 pairs. The wedge introduced during the fabrication of the microcavity allows the exciton–cavity detuning, and therefore the polariton composition, to be tuned by changing the excitation position across the sample. The high quality factor of the microcavity yields a polariton lifetime of $\approx80$ ps across the different exciton-cavity detunings used in this work. The sample is mounted in a cryostat and maintained at 4 K.

Exciton--polaritons are generated under non-resonant continuous-wave (CW) excitation using a Ti:sapphire laser. The pump is tuned to the first Bragg reflection minimum of the cavity at $\approx735$ nm. Emission from the microcavity is collected from the bottom of the lower polariton branch at $\approx774$ nm. Excitation and collection are performed in reflection geometry through the same objective in a confocal configuration. To minimize sample heating, the excitation beam is mechanically chopped at 420 Hz with a 10\% duty cycle.

The pump laser is actively stabilized both in frequency and in pointing. Frequency stabilization is implemented through feedback on the laser tuning, while beam pointing is stabilized using external piezo-actuated mirrors controlled in both real and Fourier space. The excitation profile is shaped with a spatial light modulator into a ring, producing a repulsive potential that confines polaritons in the central region. Residual reflected laser light is removed with long-pass filters, and the polariton emission is analyzed in both real and momentum space using a CCD camera and an imaging spectrometer.

Spatial coherence is measured with a Michelson interferometer incorporating a retroreflector in one arm. This geometry superimposes the condensate emission with its centrosymmetric image, allowing direct access to the first-order correlation function $g^{(1)}(\mathbf{r},-\mathbf{r})$. The 2D interference pattern (interferogram) is recorded on a CCD camera and analyzed by Fourier filtering of the first-order sideband. The spatial decay of the reconstructed coherence function is then used to extract the correlation length $\xi$.

To probe the relaxation dynamics, a weak pulsed perturbation with pulse length $\approx 100$ ps is superimposed on the continuous-wave pump. The pulse wavelength is tuned at the first minimum of the DBR stop band, as for the CW pump, except for the measurements shown in Fig.~4C, where it is set at resonance with the polariton condensate. The resulting time-dependent interference signal is detected with a streak camera, allowing reconstruction of the space- and time-resolved first-order coherence. The relaxation time $\tau$ is obtained from the decay of the spatially integrated coherence (zero delay in the interferometer) following the perturbation.

\newpage
\clearpage

\subsection*{Supplementary Text}
\

{
\renewcommand{\contentsname}{\small Contents}
\footnotesize
\renewcommand{\baselinestretch}{0.9}\selectfont
\tableofcontents
}

\newpage
\clearpage

\section{Interferometric measurement of spatial correlations and relaxation times}
\subsection{Extraction of spatial correlations.}

\begin{figure}
\centering
\includegraphics[width=0.7\textwidth]{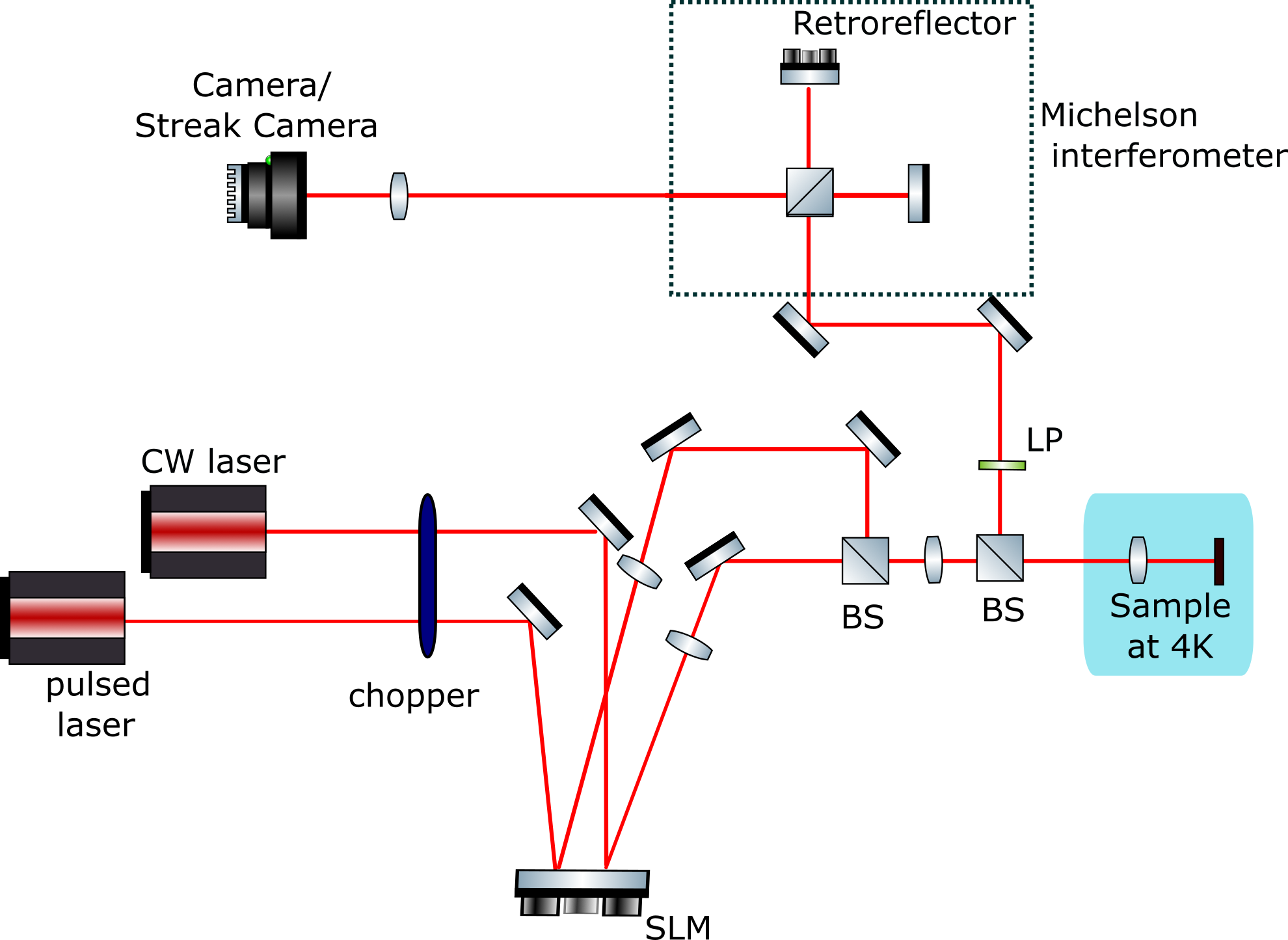}
\caption{
\textbf{Schematic of the experimental setup.} A CW Ti:Sapphire laser, shaped into a ring by an SLM, excites polaritons in a  microcavity held at 4 K cryostat. A 100-ps pulsed Ti:Sapphire laser, spatially overlapped with the CW pump, provides a non-resonant perturbation for relaxation time measurements. The photoluminescence is collected through the same objective, filtered from residual laser light with long-pass filter (LP), and analyzed in real and reciprocal space using a spectrometer, CCD and streak camera. BS stands for a beamsplitter.}
\label{fig:setup}
\end{figure}

The optical setup used in the experiments is schematically shown in Figure~\ref{fig:setup}.

The sample is held in a 4K cryostat. The polaritons are created using a continuous wave (CW) Ti-Sapphire laser with the energy blue-detuned from the lower polariton branch and set to the first Bragg reflection minimum of the cavity. 

We work in reflection geometry exciting and collecting the PL with the same objective (confocal). The excitation laser emission before the sample is chopped with the mechanical chopper at 420 Hz with 10 percent duty cycle in order to prevent the heating of the sample. We use a spatial light modulator (SLM) to shape the excitation laser in a ring shape that creates a repulsive potential for polaritons formed inside. The SLM screen is split in two parts: one controls the CW pump, while the other is used to shape the pulsed laser. We filter out the residual laser emission reflected  from the sample with long-pass filters and study the polariton emission both in real and reciprocal space using a CCD camera and imaging spectrometer. 

To characterize the coherence length $\xi$ of the condensate we utilize a Michelson interferometer with a retroreflector prism in one of the arms. The condensate emission is split in two, and one part is retroreflected. As a result, we have two copies of the condensate on the camera. The retroreflector allows us to investigate the correlations of different parts of the condensate, $r$ and $-r$, with respect to the center of the condensate. We overlap the two images on the CCD camera and obtain an interference pattern (see Figure 1 \textbf{B} in the main text) that contains information about the spatial distribution of correlations. 
The first-order coherence function $g^{(1)}(r,-r)$ is obtained from the interferometric signal by isolating the first-order sideband in Fourier space $I_1$. The extracted $g^{(1)}(r,-r)$ was verified to be robust against variations of the Fourier filtering window within a reasonable range. The central (DC) component is removed by applying the same spatial filtering procedure, allowing for a normalized reconstruction of the coherence function:

\begin{equation}
     g^{(1)}(\Delta x, \Delta y) = \frac{2|IFFT[I_1(\Delta x, \Delta y)]|}{|IFFT[I_0(\Delta x, \Delta y)]|} 
\end{equation}
The normalization procedure assumes balanced intensities in the two interferometer arms; this condition was verified experimentally by measuring the individual arm intensities and ensuring symmetric overlap. 
The profiles of $g^{(1)}(r,-r)$ are radially averaged to improve the signal to noise ratio and the correlation length is then extracted by fitting the spatial decay of $g(1) (|(r)|)$ with an exponential function $g^{(1)}(\Delta r)\propto exp(-\Delta r/\xi)$. The exponential form is expected for a finite coherence length below threshold and provides an excellent fit over the selected range; deviations from exponential behavior are within experimental uncertainty. 

\begin{figure}
\centering
\includegraphics[width=1\textwidth]{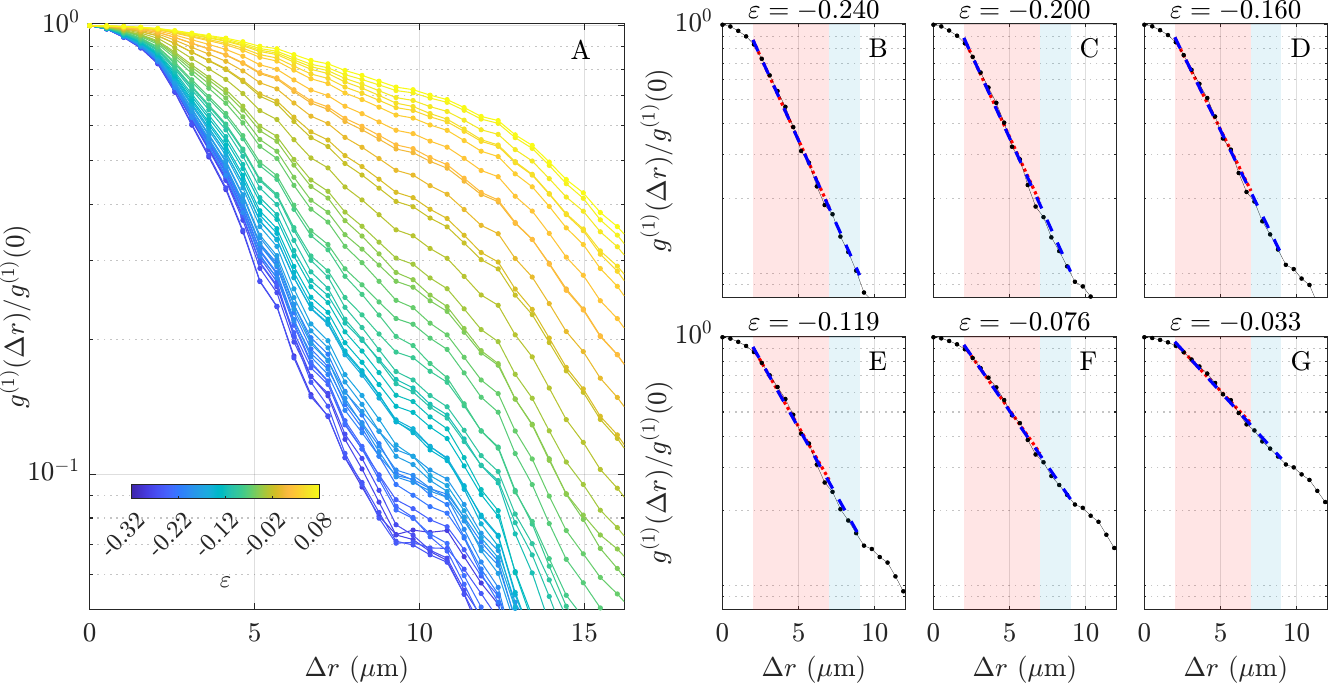}
\caption{
\textbf{Spatial decay of coherence for different pump power.} \textbf{A} $g^{(1)}(\Delta r) $ for the condensate in 15.5 $\mu m$ trap pumped at $\varepsilon = [-0.27; 0.07]$. \textbf{B} Spatial decay of $g^{(1)}(\Delta r)$ at different $\varepsilon$. The correlation length $\xi$ does not change within the fitting range $[2;8]$. Larger fitting ranges give slightly larger values of the correlation length $\xi$, while leaving the extracted static exponent $\nu$ within the experimental error.
}
\label{fig:space1}
\end{figure}

In Fig.~\ref{fig:space1}\textbf{A} we plot $g^{(1)}(\Delta r)$ for different pump intensities scanning the range corresponding to $-0.3<\varepsilon<0.1$ for a trap size of 15.5 $\mu m$. The effect of the barrier on the coherence decay is visible above 10 $\mu m$. In the range between 2-8 micron, the correlation length obtained from exponential fit is not affected by a change of the fitting range, as shown in Fig.~\ref{fig:space1}\textbf{B}. Although the finite size of the trap introduces a small structure on top of $g^{(1)}(r)$ at larger fitting ranges, the residuals remain below 3\%. Using a larger fitting range up to $11~\mu$m does not affect the extracted exponent $\nu$, although the residuals on the extracted $\xi$ increase, especially at lower powers. The stability of the exponent value is tested against different fitting ranges in space and power, giving an estimated $\nu=0.48 \pm 0.04$ for this trap size.

\subsubsection{Extraction of the relaxation time.}

To extract the relaxation time $\tau$ we add a non-resonant pulse perturbation of 100 ps to the pump. We shape the laser emission with the SLM into the ring shape and overlap it with the CW laser beam used for the excitation of the polariton condensate. The pulsed laser peak power is set to ~ 10-20 $\%$ of the CW one so that it does not affect the condensation threshold, while assuring enough signal to noise ratio to measure the dynamics of the condensate (see the next Section). 

The perturbed condensate emission follows the same optical path as in the continuous-wave (CW) experiment through the Michelson interferometer, set at zero interferometer delay, and is directed onto the slit of a streak camera with a temporal resolution of 10 ps. The relative angle between the two interferometer arms is adjusted such that the interference fringes are oriented perpendicular to the slit, enabling time-resolved detection of the interferometric signal.

This configuration allows us to monitor the temporal evolution of the coherence following the arrival of the pulse. 
The early-time dynamics can be affected by the finite response of the excitonic reservoir. In our measurements, the excitation pulse has a duration of approximately 100 ps and arrives at $t \simeq -50$ ps. Since the characteristic reservoir response is typically of the order of a few hundred picoseconds, we extract the relaxation time only from the long-time tail of the signal, starting at $t=800$ ps, corresponding to about 850 ps after the pulse arrival, well beyond the initial pulse- and reservoir-affected transient.

To extract the relaxation time $\tau$, we reconstruct the first-order coherence function $g^{(1)}(\Delta x,t)$ at zero interferometer delay as a function of the time $t$ after the pulse. The steady-state contribution under CW excitation is measured independently by blocking the pulsed laser and is subtracted from the time-resolved signal. The remaining transient is then fitted with a single exponential of the form
\[
g^{(1)}(t) \propto e^{-t/\tau},
\]
using the time window from 800 to 1600 ps. We further verified that the extracted value of $\tau$ is robust against reasonable variations of the fitting range: changing the fit window from 800--1600 ps to 1000--1600 ps does not significantly modify the resulting relaxation time. The extracted time constant $\tau$ therefore characterizes the long-time relaxation of the perturbation towards the steady state.

\begin{figure}
\centering
\includegraphics[width=1\textwidth]{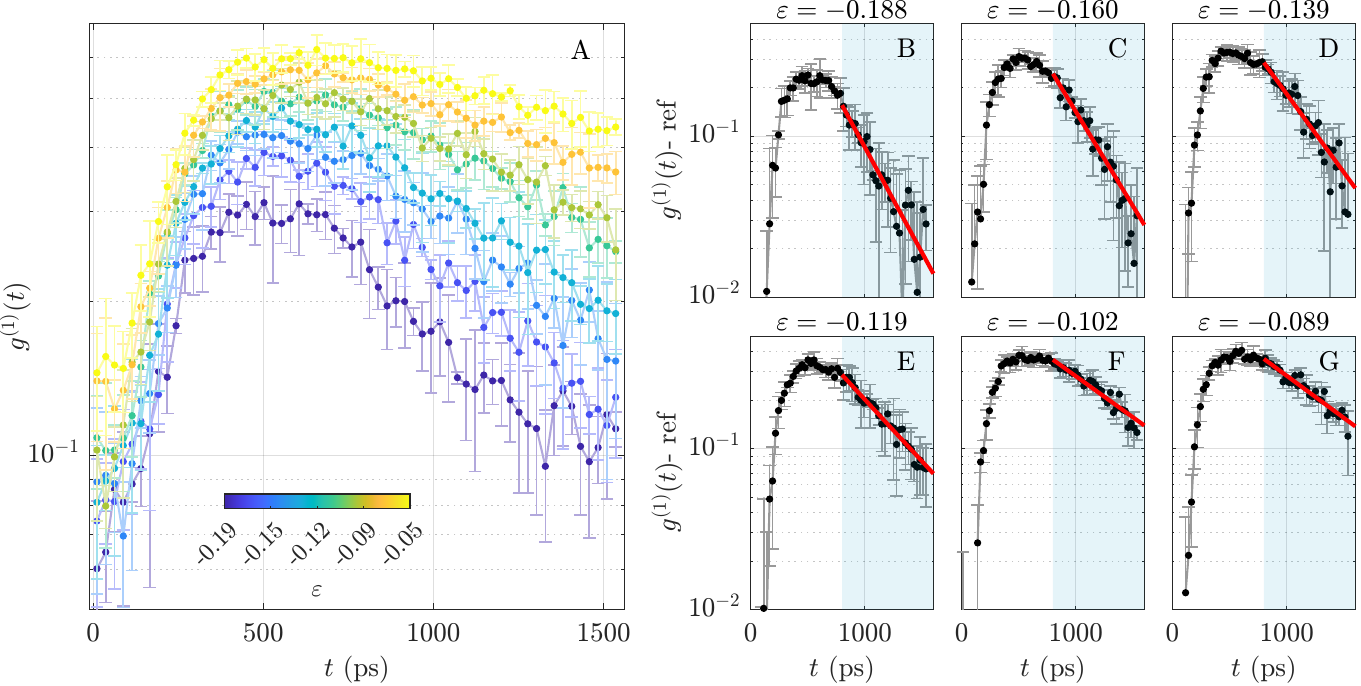}
\caption{
\textbf{Relaxation time after a perturbation } \textbf{A} First-order coherence function $g^{(1)}(\Delta x,t)$ in a 15.5 $\mu m$ trap, measured at zero interferometer delay as a function of the time $t$ after the pulse. The steady-state contribution under CW excitation is measured independently by blocking the pulsed laser and is subtracted from the time-resolved signal. \textbf{B-G} The exponential fit used to extract the relaxation time $\tau$, shown in the time window from 800 to 1600 ps.
}
\label{fig:taus}
\end{figure}

By varying the CW pump power across threshold, from $\varepsilon=-0.3$ to $\varepsilon=0.1$, while keeping the perturbation strength fixed, we extract the dependence of the relaxation time $\tau$ on $\varepsilon$. The resulting behavior exhibits a clear divergence of $\tau$ as the system approaches the condensation threshold $\varepsilon=0$ (see Fig.~\ref{fig:taus}).

\subsection{Calculating error for the experimental $\nu$ and $z$. }

To estimate the error of the measurements and the fitting for the experimentally retrieved critical exponents, we use the procedure described below.

For the static exponent $\nu$, the total error consists of three parts: the fitting error, the error associated with the uncertainty in the definition of the threshold, and the error extracted from the variation of the fitting window. To determine $\nu$, we use the dependence of the coherence length $\xi$ on the pump power $\varepsilon$ (presented, for example, in Figure 2 \textbf{A} in the main text). We fit the data with the power-law function ($\approx \varepsilon^{-\nu}$) described above. The fitting region is chosen such that the velocity v decreases linearly and the diffusion coefficient $D$ remains constant. The fit provides the value of $\nu$ together with the fitting error $e_1$.

Furthermore, we vary the threshold by $g = \pm 0.02$ by modifying the fitting function to ($\approx (\varepsilon + g)^{-\nu}$). By calculating several values of $\nu_b$ for different $b$, we determine the error $e_2$ associated with the definition of the threshold as the standard deviation of the obtained $\nu_b$ values. We also vary the fitting range by shifting it by plus - minus ten percent of the fitting window width. From the standard deviation of the resulting $\nu$ values, we obtain $e_3$, which reflects the effect of the fitting window position. The final error bar is calculated as $\sqrt{e_1^2 + e_2^2 + e_3^2}$ and is plotted e.g. in Figure 3 \textbf{E} in the main text.

We follow the same procedure to extract the error of the dynamical exponent $z$, except for the variation of the threshold, since $z$ is calculated from the dependence of $\tau$ on $\xi$ (see Figure 2\textbf{D}), which does not depend on the exact position of the threshold ($\varepsilon = 0$).

\subsection{Experimental values of $\nu$ and $z$ for different trap size. }

\begin{table}[ht]
\centering
\begin{tabular}{c c c c c c}
\hline
\hline
trap size $d$ ($\mu$m) & $|X|^2$ & $\nu$ from $\xi(\varepsilon)$ & ($\nu z$) from $\tau(\varepsilon)$ & $z = (\nu z)/\nu$ & $z$ from $\tau(\xi)$ \\
\hline
12.6 & 0.41 & $0.53 \pm 0.18$ & $0.60 \pm 0.23$ & $1.13 \pm 0.57$ & $1.18 \pm 0.06$ \\
12.9 & 0.41  & $0.46 \pm 0.09$ & $0.73 \pm 0.14$ & $1.58 \pm 0.42$ & $1.58 \pm 0.07$ \\
13.1 & 0.41  & $0.64 \pm 0.08$ & $1.21 \pm 0.19$ & $1.90 \pm 0.38$ & $1.93 \pm 0.27$ \\
14.7 & 0.41  & $0.59 \pm 0.09$ & $1.32 \pm 0.23$ & $2.22 \pm 0.50$ & $2.09 \pm 0.14$ \\
15.5 & 0.41  & $0.47 \pm 0.07$ & $1.01 \pm 0.17$ & $2.18 \pm 0.48$ & $2.17 \pm 0.15$ \\
17.5 & 0.41  & $0.53 \pm 0.08$ & $1.14 \pm 0.16$ & $2.13 \pm 0.45$ & $2.01 \pm 0.22$ \\
\hline
\hline
\end{tabular}
\caption{Critical exponents extracted from experiment. The dynamic exponent $z$ is obtained both from $(\nu z)/\nu$ and independently from $\tau(\xi)$.}
\end{table}

\section{Robustness of Critical Scaling}

The robustness of the critical scaling is tested against different experimental and numerical configurations.

\subsection{Pulse power dependence}

\begin{figure}
\centering
\includegraphics[width=0.9\textwidth]{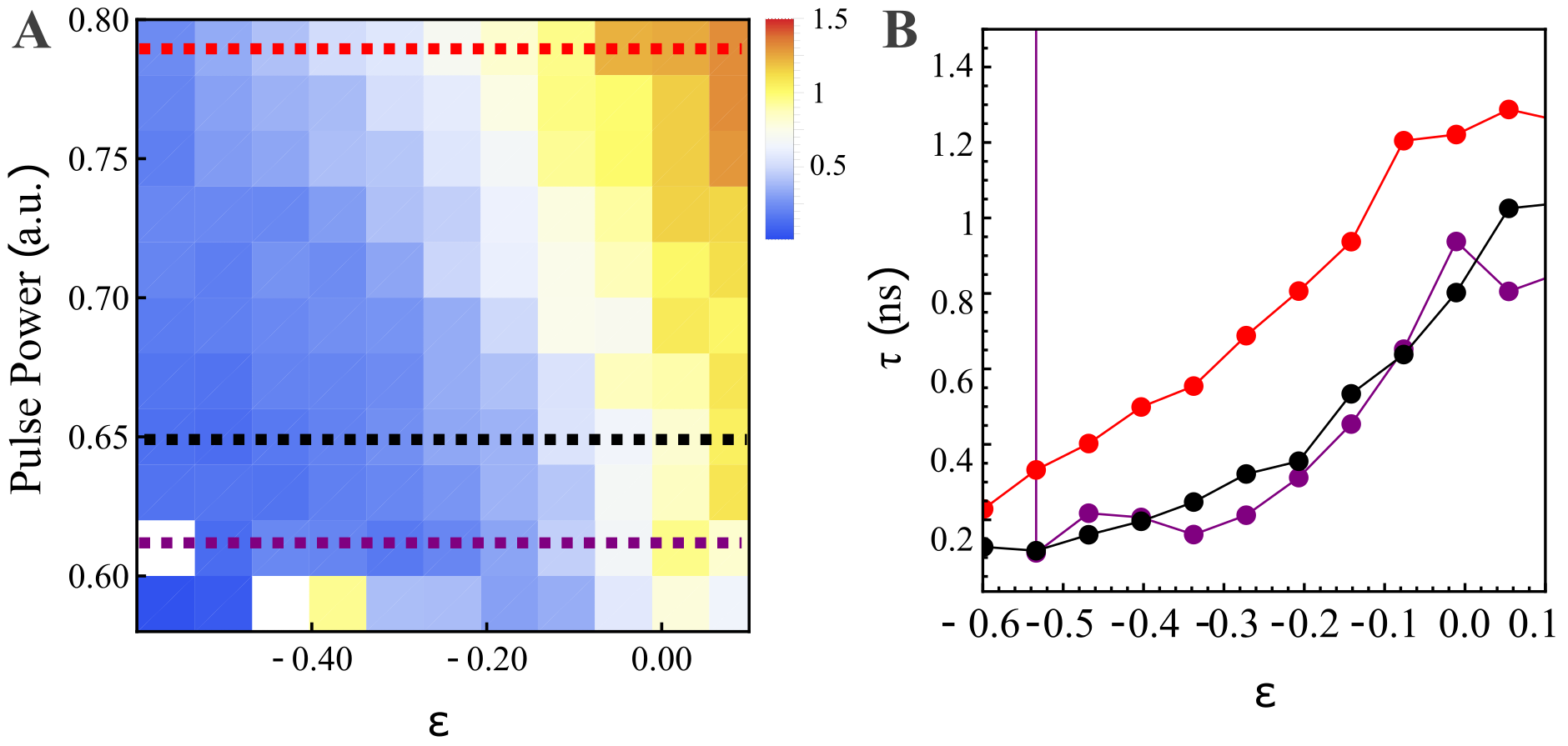}
\caption{
\textbf{Pulse power dependence.} \textbf{A} The perturbation decay time $\tau$ (color-coded) as a function of CW pump power in the units of $\varepsilon$ and pulse power in arbitrary units. \textbf{B} The dependence of the decay time $\tau$ on epsilon for the pulse power equal 0.62, 0.65  and 0.78 a.u. - purple, black and red points, respectively.
}

\label{fig:pulsepd}
\end{figure}

The pulsed excitation is used to perturb the condensate and probe its relaxation dynamics. The perturbation amplitude must therefore be chosen within a range where the extracted relaxation time is robust against changes in pulse strength. To assess this effect, we measure the coherence decay time $\tau$ for different pulse powers. Figure~\ref{fig:pulsepd}\textbf{A} shows $\tau$ as a function of both pulse power and CW pump power $\varepsilon$. For sufficiently large pulse powers, the measured decay is clearly affected by the perturbation, as evidenced by the increase of $\tau$ already for $\varepsilon<0$ shown in Fig.~\ref{fig:pulsepd}\textbf{B}. Therefore the pulse strength is kept below the regime where it modifies the observed relaxation dynamics.

\begin{figure}
\centering
\includegraphics[width=0.7\textwidth]{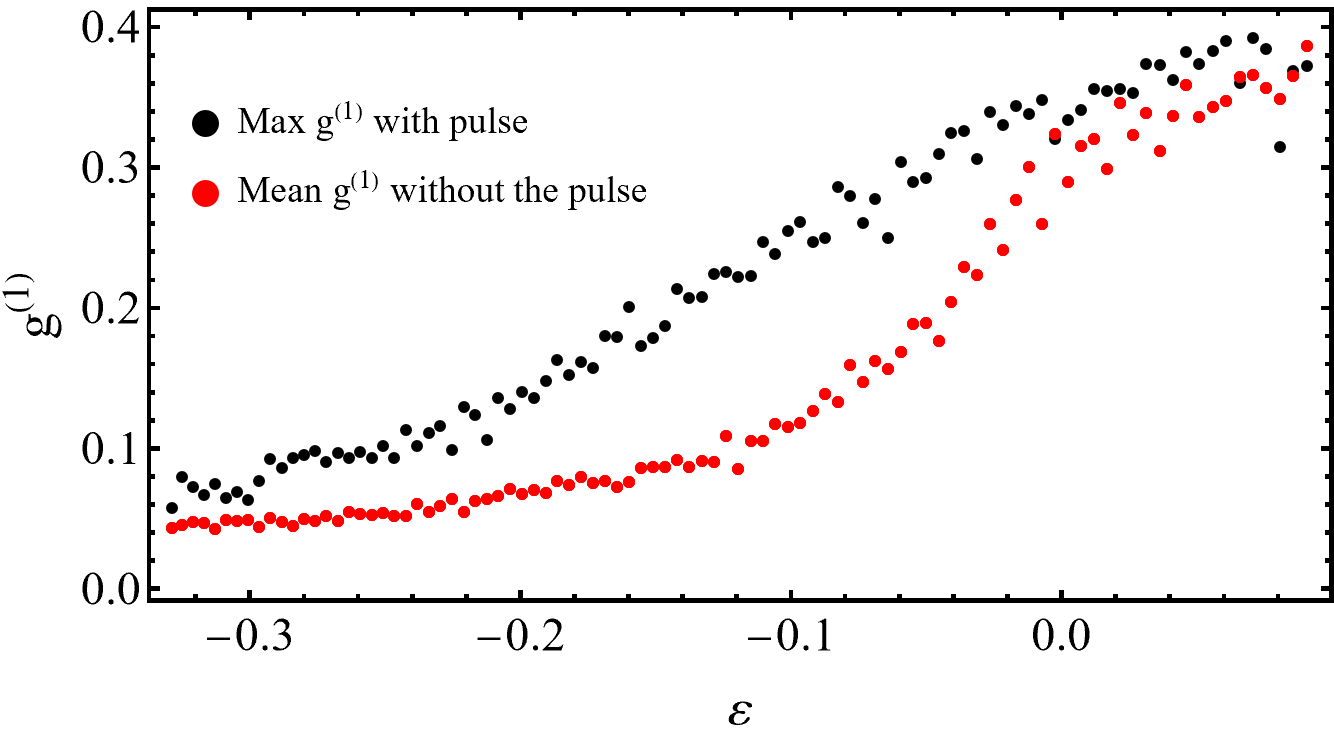}
\caption{
\textbf{Coherence of the condensate with and without the perturbation. } Red points depict the mean coherence of the condensate without the pulse as a function of $\varepsilon$. Black points depict the max $g^{(1)}$ coherence value of the condensate induced by the pulse as a function of $\varepsilon$. 
}
\label{fig:maxmin}
\end{figure}

As a second consistency check, for each experimental condition we characterize the maximum coherence induced by the pulse (Fig.~\ref{fig:maxmin}). In particular, we compare the spatially integrated coherence under CW-only excitation with the maximum integrated coherence reached after the arrival of the pulse, for the same value of $\varepsilon$. The pulse power is chosen such that the transient perturbation does not drive the system above threshold. To ensure this, we require that the maximum coherence reached after the pulse remains below the steady-state coherence measured at threshold ($\varepsilon=0$) under CW-only excitation.

At the same time, the pulse power cannot be made arbitrarily small. For too weak pulses, the induced coherence becomes comparable to the noise level at large negative $\varepsilon$, making the extraction of the relaxation time less reliable, as illustrated in Fig.~\ref{fig:pulsepd}\textbf{A}. The pulse power used throughout the main text is therefore selected as a compromise between these requirements: it must be low enough not to modify the intrinsic decay dynamics and not to transiently drive the system above threshold, yet large enough to produce a measurable perturbation over the widest possible sub-threshold range.

We also investigate numerically the effect of the perturbation strength on the system dynamics. In particular, we compute the relaxation time $\tau(\varepsilon)$ for different values of the relative excitation amplitude, defined as $\Delta \varepsilon = \Delta \mathcal{P}/\mathcal{P}_{\mathrm{th}}$. The results are shown in Fig.~\ref{fig:num_delta_P}, where we consider $\Delta \varepsilon = 0.04$, $0.08$ (the case reported in Fig.~2 of the main text), and $0.15$, corresponding to the red, blue, and yellow open circles, respectively. Within this range of excitation amplitudes, the relaxation time $\tau$ remains essentially unchanged, indicating that the extracted dynamics is robust with respect to the strength of the perturbation.

\subsection{Gaussian excitation}

To clarify the role of the excitonic reservoir in the observed dynamics, we perform complementary measurements under Gaussian excitation. In this configuration, both the CW and pulsed beams are shaped into Gaussian spots by changing the mask on the spatial light modulator. The condensate is then formed directly on top of the reservoir, maximizing reservoir--polariton overlap and interactions.

Applying the same analysis procedure as for the ring-shaped excitation, we measure the dependence of the coherence length on pump power (Fig.~\ref{fig:gauss}\textbf{A}). In this case, the increase of coherence towards threshold is much weaker than in the trapped condensate. Similarly, in the time-resolved measurements we do not observe any significant increase of the relaxation time $\tau$ on approaching threshold (Fig.~\ref{fig:gauss}\textbf{B}). Instead, $\tau$ remains approximately constant, $\tau \sim 200$ ps, over the explored range of $\varepsilon$.

The absence of any critical enhancement of $\tau$ indicates that the dynamics is governed by an approximately fixed, non-critical timescale. This behavior is consistent with reservoir-dominated relaxation: when the condensate spatially overlaps with the reservoir, the decay of the perturbation is controlled by the reservoir response rather than by the intrinsic critical slowing down of the polariton field.

In this configuration, no reliable static or dynamical critical exponent can be extracted. This comparison shows that the universal scaling reported in the main text is not a generic feature of non-resonantly driven polariton systems, but relies on the reduced reservoir overlap provided by the ring-trap geometry.

\subsection{Resonant Probe}

To exclude the possibility that the excitonic reservoir affects the observed polariton dynamics and critical behavior, we perform complementary measurements with a resonant pulse. The pulse wavelength is tuned to the condensate energy at $\varepsilon=-0.11$, so that the perturbation directly seeds the polariton field without creating an additional non-resonant reservoir. In addition, the pulsed beam is shaped into a Gaussian spot to maximize its spatial overlap with the condensate and facilitate efficient seeding.

The pulse power is kept as low as possible, so as not to modify the condensate dynamics or saturate the streak camera. Under these conditions, the resonant perturbation produces a measurable transient increase of coherence, allowing the subsequent decay to be resolved in time.

\begin{figure}
\centering
\includegraphics[width=0.5\textwidth]{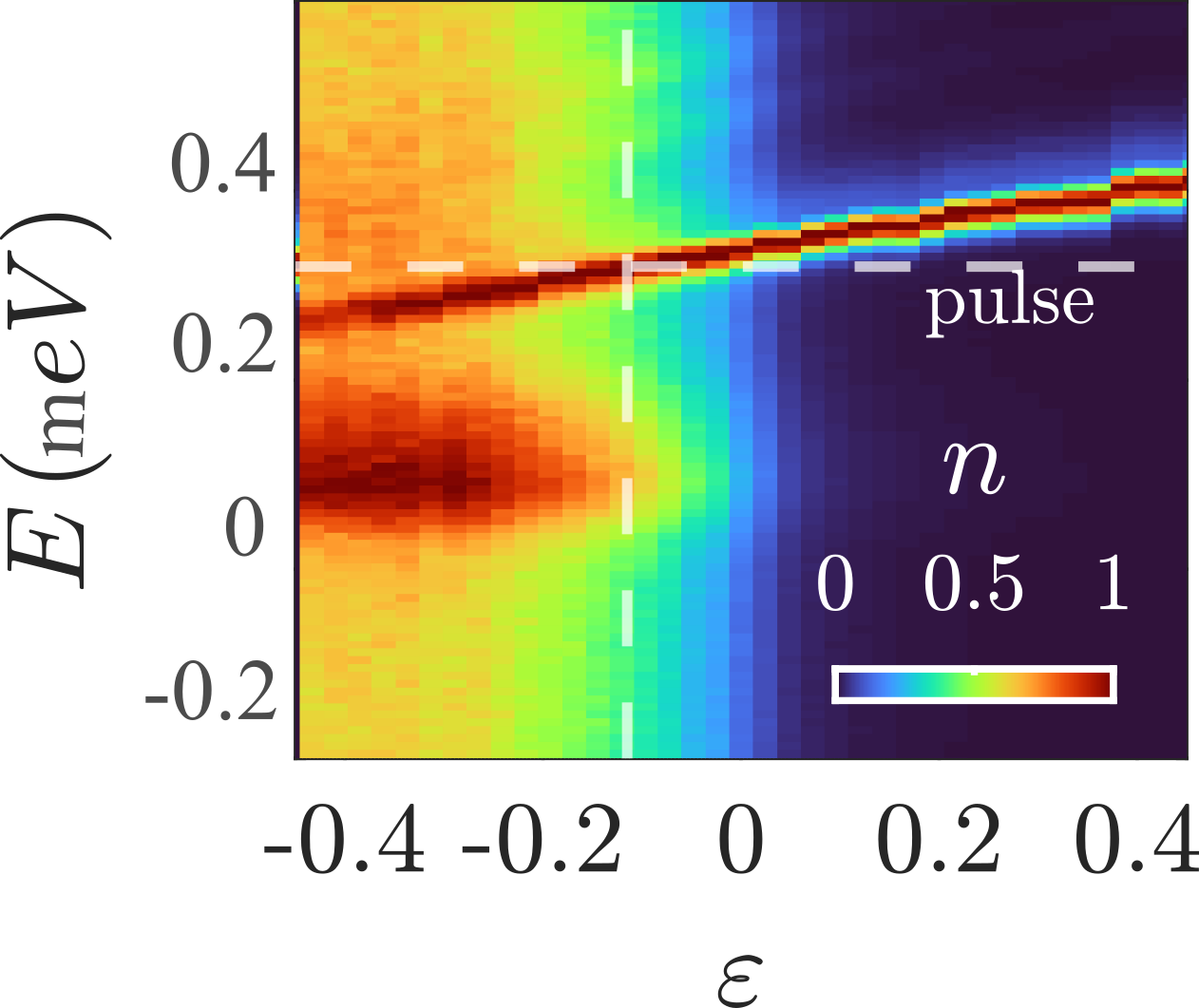}
\caption{
\textbf{Spectrum power dependence and resonant perturbation. } Spectrum power dependence of the polariton condensate in 14.7$\mu m $ diameter trap. The energy is calculated from the bottom of the lower polariton branch. The spectrum is normalized for each pump power $\varepsilon$. The dashed line shows the energy of the resonant pulsed perturbation hitting the polariton energy at approximately $\varepsilon =  -0.11$. The signal at lower energy visible for pumping powers below threshold is emission from polaritons outside the trap
}
\label{fig:resonant}
\end{figure}

\begin{figure*}
\centering
\includegraphics[width=.6\linewidth]{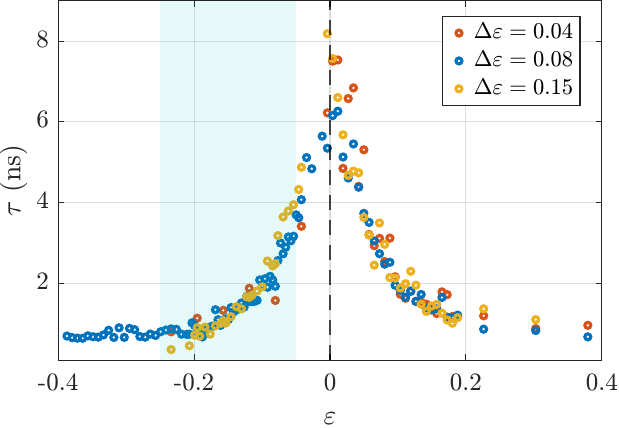}
\caption{\textbf{Dependence of the relaxation time on the Pulse Power}. The relaxation time $\tau(\varepsilon)$ is evaluated for different excitation strengths, $\Delta \varepsilon = 0.04$, $0.08$ (the case reported and discussed in the main text), and $0.15$ (red, blue, and yellow empty circles respectively). The results indicate that $\tau$ remains essentially invariant with respect to the excitation strength over the range considered in this work.}
\label{fig:num_delta_P}
\end{figure*}

\begin{figure}
\centering
\includegraphics[width=0.9\textwidth]{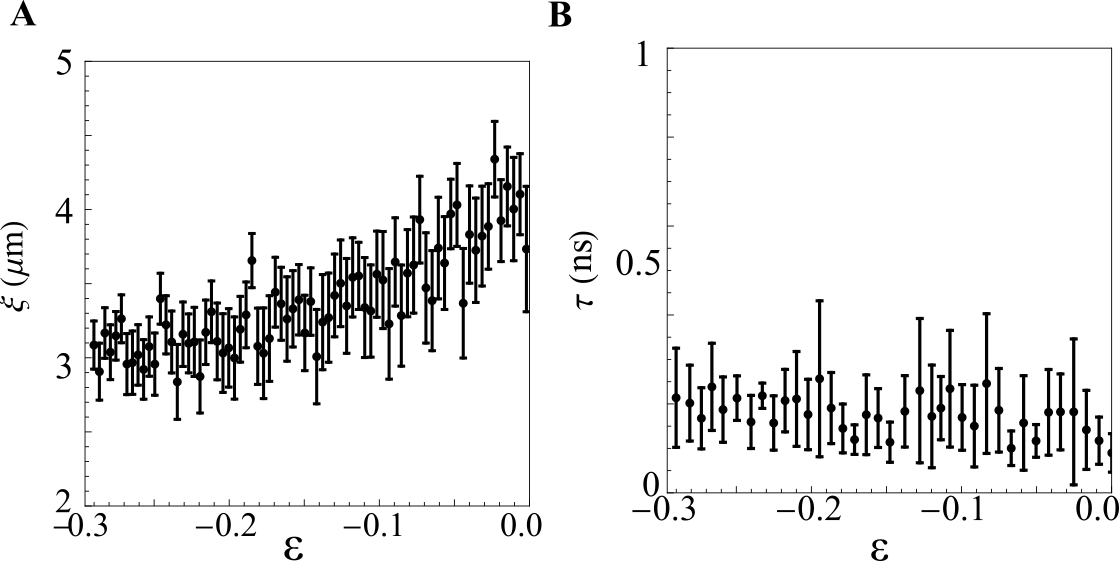}
\caption{
\textbf{Gaussian excitation.} 
\textbf{A} Coherence length $\xi$ and \textbf{B} relaxation time $\tau$ as a function of pump power $\varepsilon$ for for the condensate created with gaussian beam.
}
\label{fig:gauss}
\end{figure}

We then vary the CW pump power and extract the coherence decay time for each value of $\varepsilon$ using the same fitting procedure as in the non-resonant case. As in the main experiment, multiple measurements are collected and averaged to suppress fluctuations. The resulting dependence of $\tau$ on $\varepsilon$ is shown in Fig.~\ref{fig:fig4}\textbf{C}. Unlike the non-resonant case, the resonant measurement exhibits a maximum at $\varepsilon=-0.11$, corresponding to the pump power for which the pulse is in resonance with the condensate. For higher pump powers, the condensate energy is blue-detuned with respect to the pulse energy, reducing the effectiveness of the resonant perturbation. The normalized spectrum power dependence is shown in Figure~\ref{fig:resonant}.

Owing to the free spectral range of the pulsed laser, the pulse energy could not be tuned closer to threshold. On the low-power side of the resonance, however, the extracted scaling remains consistent with $z\simeq2$.

\subsection{Excitonic Fraction Dependence: Experiment and Numerics}

Polaritons are very flexible quasiparticles, their mass and interactions can be finely tuned by adjusting their photonic and excitonic parts. In the experiment, this is achieved by scanning the exciton–photon detuning. Here we investigate both numerically and experimentally the effect of the excitonic fraction on the observed scaling.

\subsubsection{Experiment }

Due to the fabrication process, the sample cavity is produced such that it has a small wedge from the center to the edges of the wafer. This effectively alters the energy of the photon in the cavity and thus the exciton–photon detuning. By changing the excitation spot on the sample, we can vary the detuning and thus change the photon and exciton fraction of polaritons. In the experiments, we varied the excitonic fraction from $|X|^2 = 0.41$ to $|X|^2 = 0.48$. In the numerical simulations, the longer effective polariton lifetime allows the exploration of a broader detuning range, extending the analysis up to $|X|^2 = 0.6$.

\subsubsection{Numerics: dependence on the Excitonic Fraction}

In the main text, all numerical results are presented for a fixed value of the excitonic fraction, $|X|^2 = 0.5$. To investigate the role of $|X|^2$, we repeat the numerical simulations while keeping all other parameters unchanged, varying only the excitonic fraction.

Most of the parameters entering Eq.~\eqref{eq:SCGLE} depend on the excitonic Hopfield coefficient $X$~\cite{estrecho2018single}, namely: the polariton and exciton decay rate $\gamma_\mathrm{c}, \ \gamma_\mathrm{R}$, the interaction parameters $g_\mathrm{c}, g_\mathrm{R}$, and the stimulated scattering parameter $R$. We refer to the section  ``The Stochastic Complex Ginzburg Landau Equation for a polariton system" for a detailed description of the dependence of these parameters in $|X|^2$.

\begin{figure*}
\centering
\includegraphics[width=\linewidth]{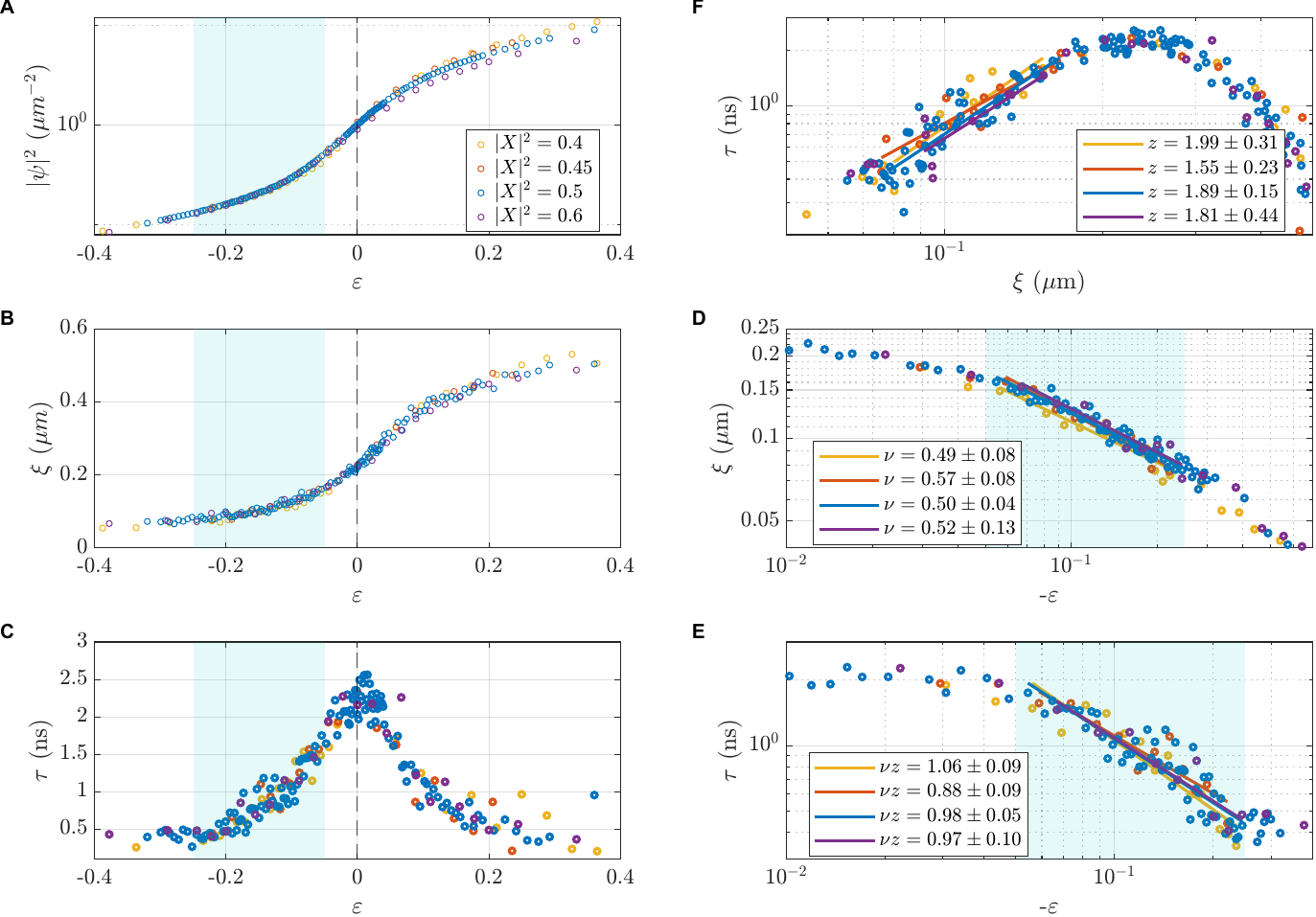}
\caption{\textbf{Numerics: Dependence on the excitonic fraction}. For different excitonic fractions, $|X|^2 = 0.4, \ 0.45, \ 0.5$ and $0.6$, the different panels report: \textbf{A)} averaged density $|\psi|^2$, as a function of $\varepsilon$ \textbf{B)} correlation length $\xi$ as a function of $\varepsilon$, \textbf{C)} relaxation time $\tau$ as a function of $\varepsilon$; \textbf{D)}, $\tau$ as a function of $\xi$, with fitting functions (solid lines). \textbf{E,F}, same as \textbf{B,C}, but in log-log scale, with fitting functions as solid lines, respectively.
Exponents and error bars are reported in the legend.  }
\label{fig:X2}
\end{figure*}

In Fig.~\ref{fig:X2}, we report the results obtained by varying $|X|^2 = 0.4$, $0.45$, $0.5$ and $0.6$, while the extracted critical exponents, reported in the legends, are shown in Fig.~4 of the main text. We observe a small variation in the density behaviour as a function of the reduced control parameter $\varepsilon$, particularly above threshold. This effect is less pronounced for the correlation length $\xi$ and the relaxation time $\tau$, where the uncertainties are larger. 
\

\section{Theoretical Model}

\subsection{{The Stochastic Complex Ginzburg Landau Equation for a polariton system}}

We describe the effective dynamics of the polariton fluid using a phenomenological model based on a stochastic complex Ginzburg--Landau equation for the two-dimensional polariton field $\psi(\mathbf{r},t)$, where $\mathbf{r} = (x,y)$ denotes the spatial coordinates and $t$ is time.
\cite{carusotto2013quantum}, which reads ($\hbar=1$)~\cite{wouters2007,chiocchetta2013,comaron2018dynamical}:
\begin{equation}
i \hbar \partial_t \psi (\textbf{r},t) = \bigg[ \left( i  \beta - 1 \right) \frac{\hbar^2 \nabla^2}{2 m_\mathrm{pol}} + g_\mathrm{c}|{\psi(\textbf{r},t)}|^2  +V(\textbf{r}) 
+ \frac{i \hbar}{2} 
\bigg(\frac{R}{\gamma_\mathrm{R}} \frac{\mathcal{P}(\textbf{r})}{1+\frac{R}{\gamma_\mathrm{R}}|{\psi}|^2 } -
\gamma_\mathrm{c} \bigg) \bigg]
\psi(\textbf{r},t) +  i \hbar dW
\label{eq:SCGLE}
\end{equation}
where $m$ is the polariton mass, $\gamma_\mathrm{c}$ and $\gamma_\mathrm{R}$ are the polariton and excitonic reservoir decay rates, $\mathcal{P}(\textbf{r})$ the external drive, $g_\mathrm{c}$ is the polariton-polariton interaction strength. $R$ is the scattering rate of reservoir particles into the condensate.
The potential exerted by high-energy excitons is modeled as a potential $V(\mathbf{r})$.
The zero-mean white Wiener noise $dW$ fulfils the condition $\langle
dW^{*}(\vec{r},t) dW (\vec{r}',t) \rangle
= 2 D_W \delta_{\vec{r},\vec{r}'}\delta_{t,t'}$. Here $D_W = [((R / \gamma_\mathrm{R}) \mathcal{P} +\gamma_\mathrm{c})/4]$, the two contributions correspond to pump-induced fluctuations and loss-induced
(vacuum) noise, respectively.
The equation above includes the phenomenological term proportional to the constant $\beta$, which quantifies the rate of energy relaxation in the system~\cite{chiocchetta2013,woutersLiew2010,wouters2010,comaron2018dynamical,zamora2020}.
The model Eq.~\eqref{eq:SCGLE} corresponds to the adiabatic approximation limit of the generalised polariton equations of motion coupled to an external reservoir~\cite{wouters2007}, which is justified once the system reaches its nonequilibrium steady state (NESS) , or if the reservoir is able to adiabatically follow the evolution of the condensate~\cite{bobrovska2015}.

\

Most of the parameters in Eq.~\eqref{eq:SCGLE} depend on the excitonic Hopfield coefficient $X$~\cite{estrecho2018single}.The latter also defines the value of the excitonic fraction $|X|^2$, and depends on the exciton–photon detuning $\Delta$ and the Rabi splitting $\hbar\Omega$ as $|X|^2~=~{1}/{2}\left(1+ {\Delta}/{\sqrt{4 \hbar^2 \Omega^2 + \Delta^2}}, \right)$.
The polariton decay rate depends on the excitonic fraction as  $\gamma_\mathrm{c} = (1-|X|^2) \gamma_\mathrm{ph}$, where $\gamma_\mathrm{ph} = 1/\tau_\mathrm{ph}$ corresponds to the inverse of the photon lifetime.
The constants $g_\mathrm{c}$ and $g_\mathrm{R}$ characterise the strengths of polariton-polariton and polariton-reservoir interactions, which become stronger for polaritons with a larger excitonic fraction. They can be estimated $g_\mathrm{c} = g_\mathrm{ex}|X|^4$, $g_\mathrm{R} = g_\mathrm{ex}|X|^2$~\cite{Bleu2020}, with $g_\mathrm{ex}$ the exciton-exciton interaction. 
Following the discussion in Ref. ~\cite{estrecho2018single,comaron2025coherence}, we assume that the stimulated scattering rate $R$ from the reservoir into the polariton states is more efficient for more excitonic polaritons, with $R = R_0 (g_\mathrm{c}/g_\mathrm{R})^2$, namely $R$ scales quadratically with $|X|^2$.

\

Above threshold the system develops a finite condensate density and spontaneously breaks the U(1) symmetry. The relevant long-wavelength degree of freedom becomes the phase $\theta$ in $\psi=\sqrt{\rho}\,e^{i\theta}$. In this regime density fluctuations are gapped, while the phase remains gapless due to the broken $U(1)$ symmetry, giving rise to a Goldstone mode. In driven--dissipative systems this Goldstone mode is diffusive at long wavelengths, or weakly propagating, as predicted theoretically~\cite{wouters2007,szymanska2006nonequilibrium} and recently observed experimentally~\cite{claude2025goldstone}.

Below threshold the system remains in the symmetric phase with vanishing order parameter in the thermodynamic limit.
Here the slow dynamics is expected to be governed by critical relaxation of the full complex order parameter $\psi$. As the transition is approached from below, the effective relaxation rate of small fluctuations vanishes and long-wavelength fluctuations become critically slow. 

\

\subsection{{Linearization to the diffusion equation}}

Below threshold, the field amplitude remains small, allowing us to linearize Eq.~\eqref{eq:SCGLE} around the symmetric state $\left< \psi \right> = 0$. Consequently, nonlinear saturation effects and inter-particle interaction terms can be neglected. 
In particular, for $|\psi|^2 \ll \gamma_\mathrm{R} / R$, one can expand
\begin{equation}
\frac{1}{1+\frac{R}{\gamma_\mathrm{R}}|\psi|^2} \simeq 1- \frac{R}{\gamma_R} |\psi|^2.
\end{equation}%
By retaining only linear terms in $\psi$, effectively neglecting saturation corrections and the interaction term $g_\mathrm{c}|\psi|^2\psi$, the equation becomes linear:
\begin{equation}
i\hbar\partial_t \psi(\mathbf r,t)
=
\Bigg[
\left(i\beta-1\right)\frac{\hbar^2\nabla^2}{2 m_{\rm pol}}
+V(\mathbf r) 
+ \frac{i \hbar}{2}\left( \frac{R}{\gamma_\mathrm{R}}
 \mathcal{P}(\mathbf r) - \gamma_\mathrm{c} \right)
\Bigg]\psi(\mathbf r,t)
+ i \hbar dW.
\label{eq:scgle_linear}
\end{equation}

Dividing Eq.~\eqref{eq:scgle_linear} by $i \hbar$ yields

\begin{equation}
\partial_t \psi(\mathbf r,t)
=
-i\left(i\beta-1\right)\frac{\hbar \nabla^2}{2 m_{\rm pol}}\psi(\mathbf r,t)
-\frac{i}{\hbar}V(\mathbf r)\psi(\mathbf r,t) 
-i\left[
\frac{i }{2} \left( \frac{R}{\gamma_\mathrm{R}}
 \mathcal{P}(\mathbf r) - \gamma_\mathrm{c} \right)
\right]\psi(\mathbf r,t)
+dW,
\end{equation}
which can be rewritten as:

\begin{equation}
\partial_t \psi(\mathbf r,t)
=
\left[
\frac{1}{2}\left(\frac{R}{\gamma_\mathrm{R}}\mathcal{P}(\mathbf r)-\gamma_\mathrm{c}\right)
+\left(\frac{\hbar \beta}{2m_{\rm pol}}+i\frac{\hbar}{2m_{\rm pol}}\right)\nabla^2  
 -\frac{i}{\hbar}V(\mathbf{r})
\right]\psi(\mathbf r,t)
+dW.
\label{eq:diffusion_psi_below_threshold}
\end{equation}
Equation~\eqref{eq:diffusion_psi_below_threshold} has the standard linear SCGLE
form
\begin{equation}
\partial_t \psi =
r(\mathbf r)\psi + (\mathcal{D}+i\mathcal{D}_i)\nabla^2\psi - \frac{i}{\hbar}V(\mathbf r)\psi + \eta,
\label{eq:linear_complex_diffusion}
\end{equation}
with coefficients
\begin{equation}
r(\mathbf r)=\frac{1}{2}\left(\frac{R}{\gamma_\mathrm{R}}\mathcal{P}(\mathbf r)-\gamma_\mathrm{c}\right),
\qquad
\mathcal{D}=\frac{\hbar \beta}{2m_{\rm pol}},
\qquad
\mathcal{D}_i=\frac{\hbar}{2m_{\rm pol}},
\qquad
\eta = dW.
\end{equation}
Below threshold one has $\mathcal{P}(\mathbf r)<\gamma_\mathrm{c} \gamma_\mathrm{R}/R$ in the pumped region, hence
$r(\mathbf r)<0$ and the field decays on average.

Eq.~\eqref{eq:linear_complex_diffusion} is interpreted as
a \emph{complex diffusion equation}: the real part controls the damping and smoothing of fluctuations.
The imaginary part determines the oscillation frequencies and wave-like propagation of Fourier modes: 
the imaginary gradient term $i\mathcal{D}_i\nabla^2\psi$ and the potential term $-iV(\mathbf r)\psi$ generate coherent phase rotation and dispersion. 

\subsubsection{Dimensionless formulation.}
Starting from the linear complex diffusive equation Eq.~\eqref{eq:linear_complex_diffusion}, it is convenient to express the dynamics in terms of the reduced pump
$\varepsilon(\mathbf r)=\mathcal{P}(\mathbf r)/\mathcal{P}_{\mathrm{th}}-1$, with
$\mathcal{P}_{\mathrm{th}}=\gamma_\mathrm{c}\gamma_\mathrm{R}/R$. This yields $r(\mathbf r)=a\,\varepsilon(\mathbf r)$,
with $a=\gamma_\mathrm{c}/2$. Since $r$ enters the equation as a
linear growth (or decay) rate, it has dimensions of inverse time, implying
that $a$ sets the intrinsic microscopic timescale. This naturally motivates
the rescaling $\tilde{t}=t\,a$ (hence, $\tilde{\tau}=\tau \,a$), which absorbs the overall rate scale into
the time derivative.
As $\mathcal{D}$ has dimensions of
length$^2$/time, this identifies the characteristic length scale
$\ell^2\sim \mathcal{D}/a$. Introducing the dimensionless coordinate
$\tilde{\xi}=\xi/\ell=\xi\sqrt{a/\mathcal{D}}$ then renders the Laplacian
term dimensionless.

As discussed in the main text, in these rescaled units the equation takes a universal form,
where the explicit dependence on microscopic rates is absorbed into the definitions of
the characteristic time and length scales. The remaining dynamics is controlled by
dimensionless parameters, such as the reduced pump $\varepsilon$ and the effective
diffusive coefficients.
In Fig.~3 (c,d) of the main text, the collapse of the dimensionless time and length
scales supports the emergence of universal diffusive behaviour.
The coefficient $a$ therefore plays a dual role: it sets the natural timescale of the
problem and, through its interplay with $\mathcal{D}$, defines the emergent length scale
$\ell = \sqrt{\mathcal{D}/a}$. At the same time, it fixes the proportionality between
the reduced pump $\varepsilon$ and the effective growth rate $r$.

\section{{Numerical Simulations}}

\subsection{{The numerical system and dynamics}}

To corroborate the experimental finding, we simulate the system by means of Eq.~\eqref{eq:SCGLE}.
The pump profile $\mathcal{P}(\mathbf{r})$ is modeled as a Gaussian function $\mathcal{P}(\mathbf{r}) = \mathcal{P}_0 \exp( -\textbf{r}^2/2 \sigma^2)$ with a width $\sigma = 3 \mu m$ chosen to match the experimental conditions.
The potential exerted by high-energy excitons is modeled as a radial potential 
$V(\mathbf{r}) \propto (r / r_{\mathrm{ring}})^{6}$, with $r_\mathrm{ring}$ the ring radius.

To reach the NESS corresponding to a given pump power, we evolve Eq.~\eqref{eq:SCGLE} from an initial noise configuration for a sufficiently long time.
Once the NESS is established, we compute the first-order correlation function $g^{(1)}(r)$. To obtain a smooth first-order correlation function, we average over both the correlators at the $x=0$ and $y=0$ positions, using the expression 
\begin{equation}
    g^{(1)}(r) \equiv [ g_x^{(1)}(r) + g_y^{(1)}(r)]/2
\end{equation} 
where 
\begin{align}
g_x^{(1)}(r) =
{\langle\psi^*_{i,0}\psi_{i+r,0}\rangle_{\mathcal{N}}}/{\sqrt{\langle |\psi_{i,0}|^2\rangle_{\mathcal{N}} \langle |\psi_{i+r,0}|^2\rangle_{\mathcal{N}}}}, \\ 
g_y^{(1)}(r) = 
{\langle\psi^*_{0,j}\psi_{0,j+r}\rangle_{\mathcal{N}}}/{\sqrt{\langle |\psi_{0,j}|^2\rangle_{\mathcal{N}} \langle |\psi_{0,j+r}|^2\rangle_{\mathcal{N}}}}.
\end{align}
Here, $\psi_{i,j} = \psi(x_i,y_j)$ and the average $\langle \dots \rangle_\mathcal{N}$ is performed over the number $\mathcal{N}$ of stochastic realisations. 
All the results presented in our study are converged with respect to the number of stochastic realisations $\mathcal{N
} = 10^3$.

\subsection{Extracting the coherence length $\xi$ and relaxation time $\tau$}

To probe the static critical properties of the system, we extract the coherence length $\xi$ from the steady-state spatial correlator of the simulated polariton field for different values of $\varepsilon$.
The spatial correlators are then fitted with the function
\begin{equation}
g^{(1)}_{\text{exp}}(r) = A_\xi e^{-r/\xi} \, ,
\label{eq:gonefitrel}
\end{equation}
where $A_\xi \sim 1$.
We choose the fitting range in $r$ sufficiently far from the trap boundaries in order to minimize the effect of the increasing potential near the trap diameter. At the same time, we exclude short distances to avoid contributions from short-range (nearest-neighbor) correlations, which do not reflect the asymptotic exponential decay used to extract $\xi$.
We repeat this procedure for several values of $\varepsilon$ below and above threshold, allowing us to extract $\xi(\varepsilon)$ in the vicinity of the critical point.
A limitation arising from the lattice discretization in our simulations is that, near the transition, the correlation length remains of the order of only a few lattice spacings ($\Delta x = 0.63\,\mathrm{\mu m}$). Therefore, the measured $\xi$ is expected to be affected by a finite-resolution background associated with the numerical grid.
We quantify this non-critical background ---and exclude it from the subsequent scaling analysis--- from the fit of $\xi$ with the function $\xi_{\Delta x} + a|\varepsilon|^{-b}$, where $\xi_{\Delta x}$, $a$, and $b$ are free parameters. In our simulations we find $\xi_{\Delta x} \approx 1\,\mathrm{\mu m}$.

\begin{figure}
\centering
\includegraphics[width=\textwidth]{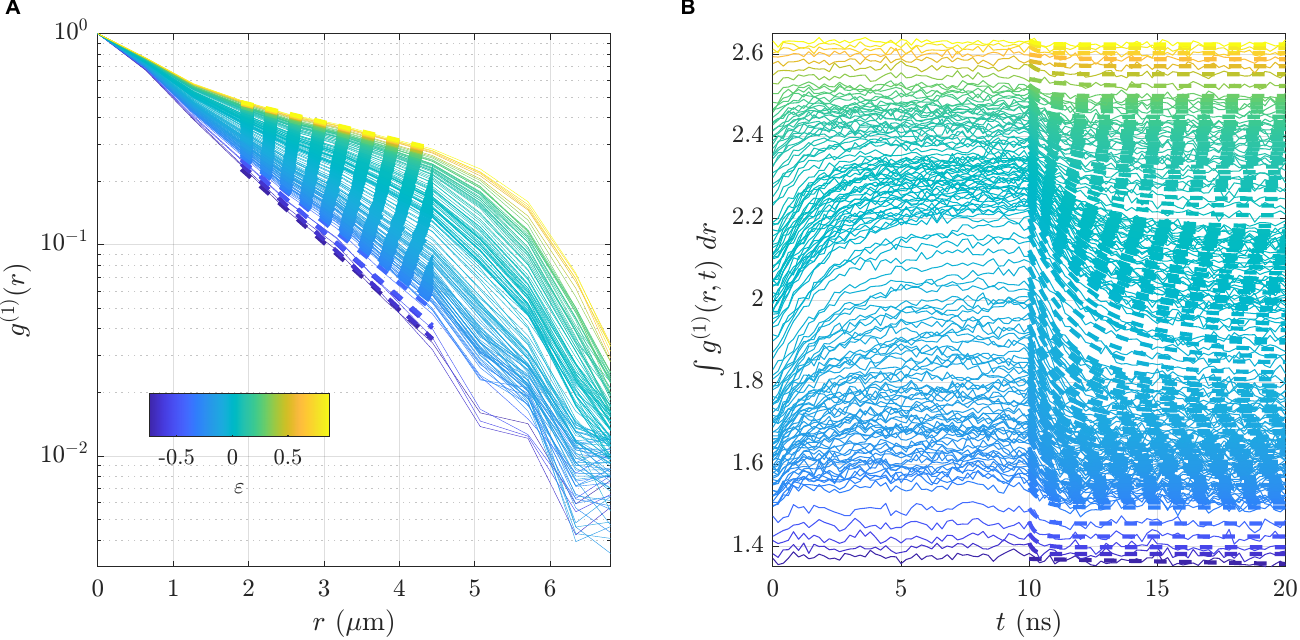}
\caption{
\textbf{Spatial and temporal decay of the first-order correlation function}. 
For the case $d = 13.6$ depicted in Fig.~2 of the main paper, we report: \textbf{a)} the spatial first-order correlation function $g^{(1)}(r)$ (solid coloured line) once the NESS is reached, for different values of $\varepsilon$, together with the corresponding exponential fits (dashed lines) according to Eq.~\eqref{eq:gonefitrel}.
\textbf{b)} the time evolution of the spatially-integrate coherence, i.e.~\eqref{eq:int_coherence}. At \(t = 0\), the system is in its non-equilibrium steady state (NESS) at a given \(\varepsilon\). We then perform a sudden quench to \(\varepsilon + \Delta \varepsilon\) and let the system evolve for \(10\,\mathrm{ns}\), allowing the new NESS to fully develop. At \(t = 10\,\mathrm{ns}\), we quench the system back to \(\varepsilon\) and monitor its relaxation toward the NESS. This second stage is fitted with a decaying exponential, from which we extract \(\tau(\varepsilon)\), reported in Fig.~2\textbf{F} of the main text. Data are reported as solid curves, the fitting functions are reported as dashed lines.
}
\label{fig:numerics2}
\end{figure}

For the case $d = 13.6\,\mathrm{\mu m}$ (shown in Fig.~2\textbf{E} of the main text), the spatial correlators together with the corresponding fitting curves are reported in Fig.~\ref{fig:numerics2}\textbf{B}.
Here the parameter $A_\xi$ ranges between $0.7$ and $1$.
Error bars on $\xi$ are obtained from the nonlinear least-squares fit as the standard error derived from the 95\% confidence interval, assuming approximately Gaussian-distributed fitting errors.

\

Following Ref.~\cite{zamora2020}, to extract the relaxation time $\tau(\varepsilon)$, we analyze the dynamical response of the system to a sudden quench of the control parameter. 
Similarly to the experimental procedure---where a pulsed excitation is applied on top of a continuous-wave (cw) background---we implement the same approach in our numerical simulations. 
Specifically, we momentarily increase the pump and then monitor the subsequent relaxation dynamics.
The system is initially prepared in the NESS corresponding to $\varepsilon$. At time $t=0$, the parameter is abruptly switched to $\varepsilon + \Delta \varepsilon$. The system is then evolved until the new NESS is reached and all fluctuations are fully developed. (in the case discussed in the main text, this occurs at $t \simeq 10\,\mathrm{ns}$). Then, the system is quenched back to $\varepsilon$. such that \begin{equation}
g^{(1)}_{\mathrm{int}}(\varepsilon + \Delta \varepsilon) 
\;\longrightarrow\; 
g^{(1)}_{\mathrm{int}}(\varepsilon) .
\end{equation}
As in Ref.s~\cite{Donner2007Science,Hadzibabic2006} and as implemented in the experiments discussed in this work,  we then monitor the subsequent relaxation of the space-integrated equal-time first-order correlation function evaluated at time $t$
\begin{equation}
g^{(1)}_{\mathrm{int}}(t) = \int g^{(1)}(r,t)\, dr \, ,
\label{eq:int_coherence}
\end{equation}
with 
\begin{equation}
g^{(1)}(r,t) = 
\frac{\left\langle \psi^*(\mathbf{x},t)\,\psi(\mathbf{x}+\mathbf{r},t) \right\rangle}
{\left\langle |\psi(\mathbf{x},t)|^2 \right\rangle} \,,
\end{equation}
which captures the global build-up of coherence in the system.

Following the quench, $g^{(1)}_{\mathrm{int}}(t)$ relaxes towards its new steady-state value. 
Then, for each value of $\varepsilon$, the relaxation curve $g^{(1)}_{\mathrm{int}}(\varepsilon,t)$ is fitted with a single-exponential form
\begin{equation}
A_\tau + B_\tau e^{-t/\tau} \, ,
\end{equation}
where $\tau$ defines the characteristic relaxation time and $A_\tau$ corresponds to the asymptotic steady-state value.
The parameter $A_\tau$ is determined by averaging $g^{(1)}_{\mathrm{int}}(t)$ over the time interval in which the nonequilibrium steady state (NESS) has been reached.

Repeating this procedure for different $\varepsilon$ allows us to extract the dependence $\tau(\varepsilon)$ in the vicinity of the critical point.
For the case shown in Fig.~2\textbf{F} of the main text ($d=13,6 \ \mathrm{\mu m}$), the spatially-integrated equal-time correlators together with the corresponding fitting curves are reported in Fig.~\ref{fig:numerics2}\textbf{B}. Invariance of the extracted $\tau(\varepsilon)$ from the choice of $ \Delta \varepsilon =  \Delta \mathcal{P}/\mathcal{P}_\mathrm{th}$ is discussed in the section ``Pulse power dependence'', above. We note that the parameter $B_\tau$ increases approximately linearly as the system approaches the threshold, taking values in the range $0 < B_\tau < 0.3$, and exhibiting a maximum at the threshold point.
Error bars on $\tau$ are obtained from the nonlinear least-squares fit as the standard error derived from the 95\% confidence interval, assuming approximately Gaussian-distributed fitting errors.

\

\subsection{{Numerical main results: different trap sizes and in the uniform limit}
}

To investigate critical dynamics,  we porform simulations with varying system sizes. In addition, we consider the case in which the trap is removed entirely and solve the model in a spatially uniform geometry with periodic boundary conditions (PBC).

In Fig.~\ref{fig:finite_size_sims} we report the extracted density, coherence length $\xi$, relaxation time $\tau$, diffusive coefficient $\mathcal{D}$, and critical slowing-down ratio $v$ in panels (a–e), respectively, as functions of the reduced control parameter $\varepsilon$. For the trapped configurations we consider three diameters, $d = 13.6$, $27.2$, and $54.4~\mathrm{\mu m}$. In the uniform case with PBC, the system is simulated on a square lattice of size $L = 31.7~\mathrm{\mu m}$.

Panel \textbf{A} shows the steady-state polariton density as a function of $\varepsilon$. As expected, the density increases as the system undergoes the transition from the disordered to the ordered phase. A detailed discussion of the determination of the critical point and its dependence on system size is provided above. The solid lines correspond to the fits described in Eq.~\eqref{eq:th_fit}. In all cases, the transition occurs at a polariton density of approximately $1~\mathrm{\mu m}^{-2}$, in agreement with the literature~\cite{Alnatah2024}.

Panel \textbf{B} displays the coherence length $\xi$ as a function of $\varepsilon$. For a fixed pump strength, $\xi$ increases as the trap size is enlarged, consistent with reduced finite-size constraints. Importantly, at the critical point the coherence length remains well below the system size in all trapped cases, which is a necessary condition for a meaningful analysis of critical scaling. 

In the uniform PBC configuration, the divergence of $\xi$ near the critical point is more pronounced, reflecting the reduced influence of finite-size effects. The solid lines represent a guide to the eye.

In panel \textbf{C} we report the relaxation time $\tau$. Similarly to the coherence length, $\tau$ increases with system size for a given pump strength. However, in contrast to $\xi$, the behavior of $\tau$ is approximately symmetric with respect to the critical point. The solid lines correspond to the fitting functions discussed in the sections above and in the main text, evaluated within the fitting region. Close to the transition, a clear saturation of $\tau$ is observed when approaching the critical point from both sides, which we attribute to finite-size effects.

Panel \textbf{D} shows the effective diffusive coefficient,
\begin{equation}
\mathcal{D} = \frac{\xi^2}{\tau}.
\end{equation}
Remarkably, $\mathcal{D}$ exhibits a pronounced plateau within the fitting region, supporting the diffusive nature of the critical dynamics. The value of $\mathcal{D}$ extracted from this flat region provides a numerical estimate of the effective diffusion constant, which will be used in the scaling analysis presented in the following sections.

We observe that $\mathcal{D}$ increases with system size. Physically, this is expected since a less confined condensate behaves as a more freely diffusing fluid. Notably, the numerically extracted diffusion coefficient does not coincide with the value obtained from the linearized theory,
\begin{equation}
\mathcal{D}_\mathrm{lin} = \frac{\hbar \beta}{2 m_\mathrm{pol}},
\label{eq:D_lin}
\end{equation}
where $m_\mathrm{pol}$ is the polariton effective mass and $\beta$ the relaxation parameter. 
For the parameters used in the simulations, $\mathcal{D}_\mathrm{lin} = 16.05 \ \mathrm{\mu m ^2 / ns}$.
This discrepancy indicates that nonlinearities and finite-size or boundary effects renormalize the effective diffusion constant away from its bare linear-theory prediction. Nevertheless, as the system size increases, the numerical value of $\mathcal{D}$ progressively approaches the linear-theory estimate.

Finally, in panel \textbf{E} we plot the ratio
\begin{equation}
v = \frac{\xi}{\tau},
\end{equation}
which characterizes the effective propagation velocity of fluctuations and quantifies critical slowing down. As the threshold is approached from below, $\mathrm{v}$ decreases, reflecting the progressive slowing of fluctuations near criticality, and increases again in the ordered phase. 

Within the fitting region, $\mathrm{v}$ decreases from approximately $2$ to $0.5~\mathrm{\mu m/ns}$, displaying only a weak dependence on system size. This behavior further supports the diffusive scaling scenario discussed in the main text.

\

\subsection{Spatial and temporal scaling: fitting procedure and error determination}

We now discuss the fitting criteria and the determination of errors in the universal region of the critical slowing down.
Each data point for $\xi$ and $\tau$ has a statistical error arising from averaging over many realizations and from the adopted fitting procedure; the calculation of the error is described in the relevant aforementioned sections.

We fit such data with the appropriate power law curve.
Specifically, the spatial coherence approaching the critical point is fitted with
\begin{equation}
\xi(\varepsilon) = A^\xi_{\varepsilon} |\varepsilon|^{-\nu},
\end{equation}
where $A_{\varepsilon}$ and $\nu$ are free parameters, with $A^\xi_{\varepsilon}$ a non-universal constant, 
and $\nu$ denoting the static critical exponent.
The relaxation time $\tau$ as a function
of the control parameter in the disordered phase are instead fitted with the function 
\begin{equation}
\tau(\varepsilon) = A^\tau_{\varepsilon} |\varepsilon|^{-\nu z}.
\end{equation}
with $A^\tau_{\varepsilon}$ a non-universal constant.

In Fig.~\ref{fig:finite_size_sims_scaling_fin_size_effects} (a,c), we report the dimensionless quantities 
$\xi \sqrt{a / \mathcal{D}}$ and $\tau a$ as functions of $\varepsilon$ for different trap sizes and for the periodic boundary condition (PBC) case, shown in log--log scale. 
Here $a = \gamma_\mathrm{c} / 2$, where $\gamma_\mathrm{c}$ the polariton loss rate.
As the quantities $\xi$ and $\tau$ may display different universality regions, when analysed independently as functions of $\varepsilon$, we determine the interval of $\varepsilon$ in which $v(\varepsilon)$ exhibits a clear power-law behaviour and use this interval as the fitting window for the fits of $\xi(\varepsilon)$, $\tau(\varepsilon)$, and $\tau(\xi)$.
The fitted curves are shown as solid lines while the extracted exponents and their uncertainties are reported in the legend.

We extract the exponents with error bars arising from the fitting procedure, where the uncertainty of each data point in the input series is taken into account.
The uncertainty on the critical exponents is obtained from the 95\% confidence interval returned by the nonlinear least-squares fit. The standard error is estimated as half the width of the confidence interval divided by 1.96, assuming approximately Gaussian-distributed fitting errors.
The variation of the threshold is not included in the error propagation, as the threshold is determined with high precision.
To account for additional uncertainty in identifying the universal scaling region, and following the same approach used in the experimental error analysis, we vary the fitting interval by 10\% inward from both the lower and upper bounds. The scaling exponents are then recalculated over these modified intervals, and the uncertainty is estimated as the standard deviation of the resulting values of  $\nu$.

In Fig.~\ref{fig:FIG_SM_z_tau_xi}, we reproduce Fig.~3\textbf{D} of the main text, namely $\tau(\xi)$ for the different trap sizes investigated, including the fitting curves and the additional case with periodic boundary conditions.
In the main text, the dynamical exponent $z$ and its associated error bars reported in Fig.~3\textbf{F} are obtained as $z = (\nu z)/\nu$, where $(\nu z)$ is extracted from the fit of $\tau(\varepsilon)$, and $\nu$ from the scaling of $\xi(\varepsilon)$. The uncertainty on $z$ is evaluated through standard error propagation, combining the independent uncertainties on $(\nu z)$ and $\nu$. We choose to report this estimate rather than the value obtained from fitting $\tau(\xi)$, as the former is affected by smaller uncertainties. Nevertheless, the value of $z$ extracted from the alternative method is consistent, within error bars, with the estimate obtained from $z = (\nu z)/\nu$.
All exponents extacted are collected in Tab.~\ref{tab:critical_exponents}.

\begin{table}[ht]
\centering
\begin{tabular}{c c c c c c}
\hline
\hline
trap size $d$ & $|X|^2$ & $\nu$ from $\xi(\varepsilon)$ & ($\nu z$) from $\tau(\varepsilon)$ & $z = (\nu z) / \nu$ & $z$ from $\tau(\xi)$ \\
\hline
$13.6$ $\mu$m & 0.5 &  $0.50 \pm 0.04$ & $0.98 \pm 0.06$ & $1.96 \pm 0.09$ & $1.89 \pm 0.15$ \\
$27.2$ $\mu$m & 0.5& $0.55 \pm 0.03$ & $1.09 \pm 0.08$  & $1.98 \pm 0.09$ & $2.03 \pm 0.10$ \\
$54.4$ $\mu$m & 0.5& $0.54 \pm 0.02$ & $1.07 \pm 0.21$  & $1.98 \pm 0.20$ & $2.2 \pm 0.22$ \\
$31.7$ $\mu$m (PBC) & 0.5 & $0.56 \pm 0.11$ & $1.06 \pm 0.02$ &  $1.88 \pm 0.19$ & $1.86 \pm 0.29$ \\
\hline
$13.6$ $\mu$m & 0.4 &  $0.49 \pm 0.08$ & $1.06 \pm 0.09$ & $2.16 \pm 0.20$ & $1.99 \pm 0.31$ \\
$13.6$ $\mu$m & 0.45 &  $0.57 \pm 0.08$ & $0.88 \pm 0.09$ & $1.71 \pm 0.14$ & $1.55 \pm 0.23$ \\
$13.6$ $\mu$m & 0.6 &  $0.52 \pm 0.13$ & $0.97 \pm 0.10$ & $1.87 \pm 0.25$ & $1.81 \pm 0.44$ \\
\hline
\hline
\end{tabular}
\caption{Critical exponents extracted from simulations. The dynamic exponent $z$ is obtained both from $(\nu z) / \nu$ and independently from $\tau(\xi)$.}
\label{tab:critical_exponents}
\end{table}

\

\subsection{Trap-size scaling analysis.}

Finite-size scaling theory plays a crucial role in understanding critical phenomena in systems of limited size~\cite{Tauber_2014}. True thermodynamic singularities arise only in the thermodynamic limit, where the number of degrees of freedom tends to infinity while densities of extensive quantities remain fixed. However, numerical simulations and experiments are always performed on finite or confined systems, making it necessary to understand how observables approach their infinite-system behavior as the system size increases. 
Near a critical point, the correlation length $\xi$ of the infinite system grows and eventually becomes comparable to the system size $L$, introducing an additional relevant length scale. When $\xi \ll L$, the system behaves similarly to the infinite system and critical singularities can be observed. In contrast, when $\xi \sim L$, finite-size effects become important and the singular behavior is rounded. These considerations lead to the finite-size scaling framework, where thermodynamic quantities are described by scaling functions that depend on the ratio $\xi/L$.

\begin{figure}
\centering
\includegraphics[width=.7\linewidth]{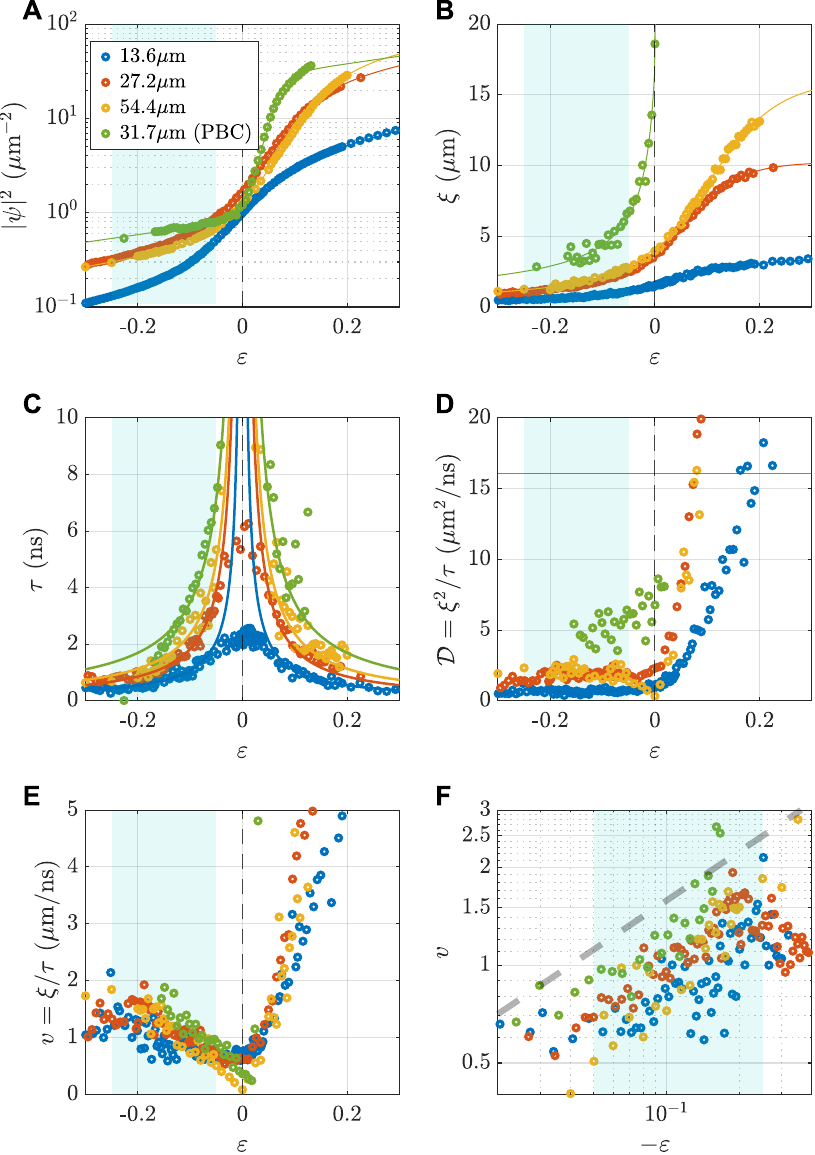}
\caption{\textbf{Numerical simulations}. The polariton density \textbf{\textbf{A}}, coherence length $\xi$ \textbf{(B)}, relaxation time $\tau$ \textbf{(C)}, diffusive coefficient $\mathcal{D}$ \textbf{(D)}, and critical slowing-down ratio $v$ \textbf{(E)}, are plotted as functions of the reduced control parameter $\varepsilon$, for three different trap sizes and for the periodic boundary condition case. In panel \textbf{D}, the red solid line indicate the value $\mathcal{D}_\mathrm{lin}$, see Eq.~\eqref{eq:D_lin}. Panel \textbf{F} depicts the critical slowing down rate $v$ in log-log scale. Grey dashed line corresponds to $v \propto \varepsilon^{\nu(z-1)}$, assuming $\nu=0.54$ and $z=2.02$.
}
\label{fig:finite_size_sims}
\end{figure}

In the simulations discussed in this work, we find that within the fitting region the condition $\xi \ll d \equiv L$ is well satisfied (see Fig.~\ref{fig:finite_size_sims}\textbf{C}). Still, for systems subject to an external trapping potential it has been predicted that critical scaling is modified by the spatial inhomogeneity~\cite{campostrinivicari2009}. 
In particular, when the system is subjected to a trapping potential of the form $V(r)\propto (r/d)^p$, the trap size $d$ plays a role analogous to the system size in finite-size scaling.

The corresponding framework~\cite{campostrinivicari2009} predicts exact universal scaling forms for observables near the critical point.
Within trap-size scaling, the correlation length $\xi$ and the characteristic relaxation time $\tau$ obey generalized scaling relations
\begin{equation}
\xi(d,\varepsilon) = d^{\theta} f\!\left(\varepsilon\, d^{\theta/\nu}\right),
\qquad
\tau(d,\varepsilon) = d^{z\theta} g\!\left(\varepsilon\, d^{\theta/\nu}\right),
\end{equation}
where $\varepsilon$ denotes the distance from the critical point, $\nu$ is the correlation-length exponent, $z$ the dynamical critical exponent, and $\theta$ the trap exponent, which depends on the power $p$ of the confining potential. The functions $f$ and $g$ are universal scaling functions.

\begin{figure*}
\centering
\includegraphics[width=\linewidth]{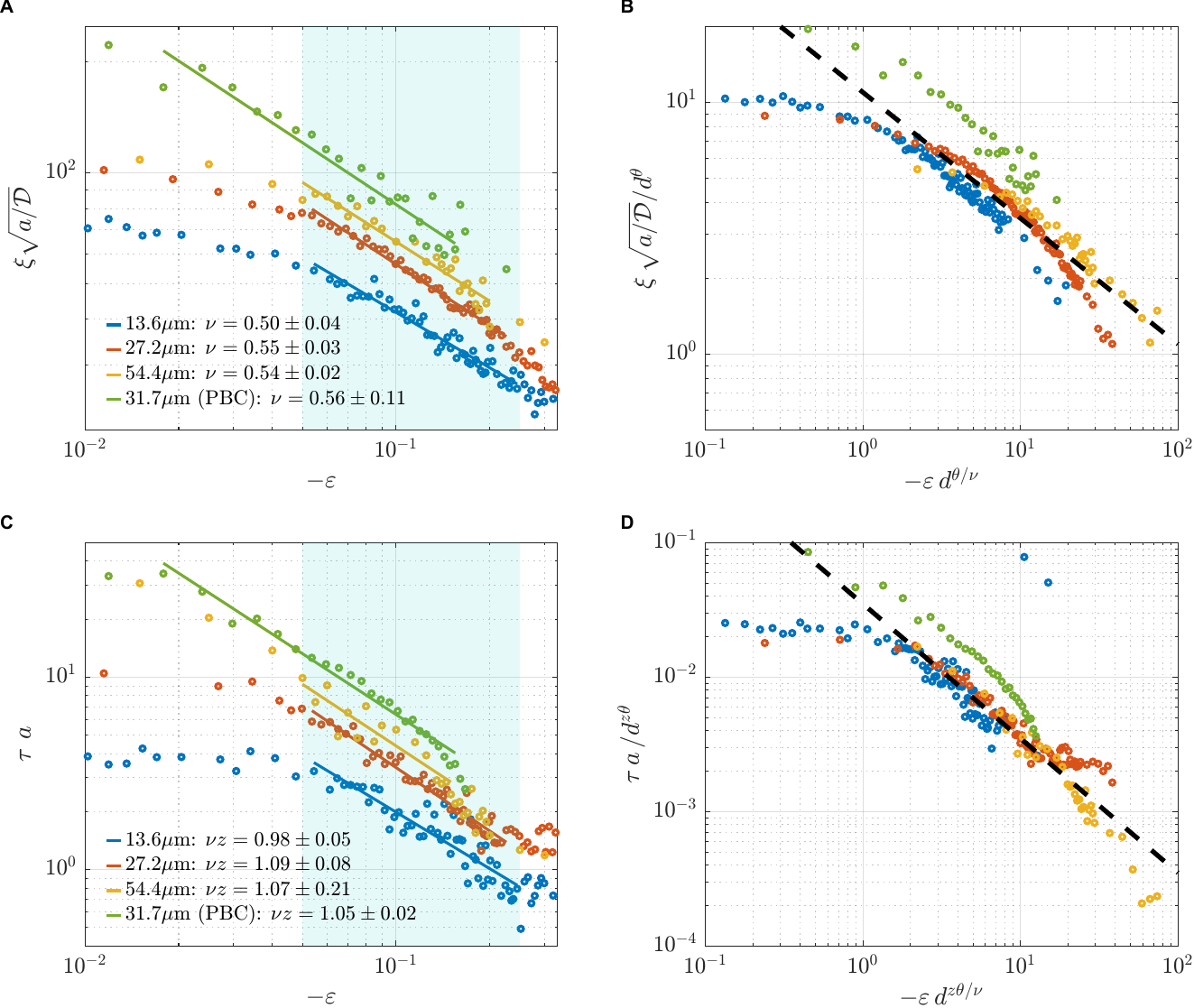}
\caption{\textbf{Numerical scaling and finite size effects}.
\textbf{A} Correlation length $\xi\sqrt{a / \mathcal{D}}$ and \textbf{C} relaxation time $\tau a$ as functions of the distance from the critical point $\varepsilon$, shown in log--log scale for different trap sizes $d$ and for periodic boundary conditions (PBC). Solid lines are power-law fits within the universal scaling region, with fitting intervals determined from the power-law behaviour of the critical slowing-down rate $v=\xi/\tau$. Extracted critical exponents and their uncertainties are reported in the legend.
\textbf{B} Data collapse of the rescaled correlation length $\xi \sqrt{a/D}  / d^{\theta}$ as a function of $\varepsilon d^{\theta/\nu}$, demonstrating trap-size scaling.
\textbf{D} Data collapse of the rescaled relaxation time $\tau a / d^{z\theta}$ versus $\varepsilon d^{z \theta/\nu}$. The collapse is obtained using $\theta=0.75$, $\nu=0.55$, and $z=2.3$. As expected, data for PBC do not collapse onto the same universal curve but retain the expected power-law behaviour.
}
\label{fig:finite_size_sims_scaling_fin_size_effects}
\end{figure*}

\begin{figure*}
\centering
\includegraphics[width=.65\linewidth]{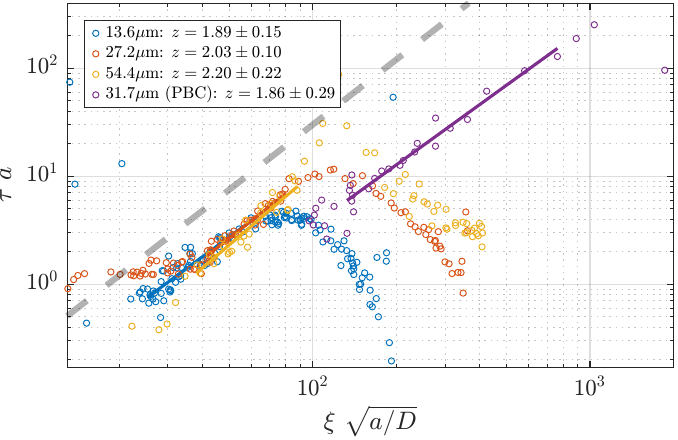}
\caption{\textbf{Numerical scaling of the dependence of $\tau$ on $\xi$}.
The relaxation time $\tau a$ as a function of the correlation length $\xi\sqrt{a / \mathcal{D}}$ is depicted in log--log scale for different trap sizes $d$ and for periodic boundary conditions (PBC). Solid lines are power-law fits within the universal scaling region. Extracted critical exponents and their uncertainties are reported in the legend. The dashed gray line corresponds to $z = 2.02$.
}
\label{fig:FIG_SM_z_tau_xi}
\end{figure*}

In our analysis we test these predictions by performing a data collapse for different trap sizes $d$. The correlation length is first expressed in diffusion units by dividing by the microscopic scale $\ell=\sqrt{\mathcal{D}/a}$, where $a$ and $\mathcal{D}$ are the characteristic rate and diffusion parameters of the model. According to the scaling form above, the correlation-length data should collapse when plotting
${\xi}/{\ell\, d^{\theta}}$ as a function of $ \varepsilon\, d^{\theta/\nu}$.
This procedure is implemented in Fig.~\ref{fig:finite_size_sims_scaling_fin_size_effects}\textbf{B}, where data obtained for different trap sizes collapse onto a single curve when expressed in these rescaled variables.

An analogous analysis is performed for the relaxation time $\tau$. For diffusive dynamics the dynamical exponent is $z$, so the trap-size scaling prediction reads
\begin{equation}
\tau(d,\varepsilon) = d^{z\theta} g\!\left(\varepsilon\, d^{\theta/\nu}\right).
\end{equation}
Therefore, the appropriate collapse variables are
${\tau}a/{d^{z\theta}}$ vs $\varepsilon\, d^{ \theta/\nu}$,
which we plot in Fig.~\ref{fig:finite_size_sims_scaling_fin_size_effects}\textbf{D}. 

The trap exponent $\theta$ is varied to optimize the data collapse across different trap sizes. According to Ref.~\cite{campostrinivicari2009}, for a trapping potential scaling as $V(r)\propto r^{6}$ the predicted trap exponent is $\theta = 0.75$. Using this value, together with the critical exponents $\nu=0.55$ and $z=2.3$ extracted in the main text, we obtain a good collapse of the data, as shown in Fig.~\ref{fig:finite_size_sims_scaling_fin_size_effects}(\textbf{B},\textbf{D}). This behavior is consistent with the expected critical scaling of the relaxation dynamics in the presence of the trap. As anticipated, the simulations performed with periodic boundary conditions (PBC) do not collapse onto the same curve, although they still display the expected power-law scaling.

We note, however, that increasing $\theta$ up to $\theta \approx 1$ does not significantly modify the apparent scaling, due to the relatively large uncertainties in the data. In contrast, reducing $\theta$ toward $0$ clearly deteriorates the collapse, indicating that a finite trap exponent is required to describe the observed scaling behavior.

\subsubsection{Parameters and further numerical details}
In order to match the experimental measurements, we use the following parameters, consistent with previous simulations of similar samples~\cite{panico2023,Alnatah2024,comaron2025coherence}:
$m=3.6 \times 10^{-5}~m_e$ with $m_e$ the electron mass, $\tau_\mathrm{ph} = 135~{ps}$, 
$\gamma_\mathrm{R} = 10^{-3}~\mathrm{ps}^{-1}$, 
$g_\mathrm{ex}~=1.12~{\mu eV \mu m^2}$, 
$R_0 = 1.4 \times 10^{-4}~{\mu m^2 ps^{-1}}$. 
For the relaxation coefficients we use $\beta = 0.01$. 
If not stated otherwise: the pump ring radius is $r_\mathrm{ring} = 6.8 {\mu m}$, $\Delta \varepsilon = 0.08$, the excitonic fraction is $|X|^2 = 0.5$.
We simulate the dynamics of the polariton system by numerically integrating the stochastic differential equations Eq.~\eqref{eq:SCGLE} for the polariton field.
The numerical integration is performed on a two-dimensional lattice with Dirichlet boundary conditions and lattice spacing $\Delta x = 0.63\mu m$. 
The integration of Eq.~\eqref{eq:SCGLE} in time is performed by using the XMDS2 software framework~\cite{dennis2013xmds2}. Specifically, we use a fixed time step which ensures stochastic noise consistency, and a fourth-order Runge-Kutta algorithm. The complex term $dW$ has been implemented using the Wiener noise. We use a double precision SIMD-oriented Fast Mersenne Twister (dSFMT) algorithm to generate the random variable.

\

\section{{Identification of the critical point}}

\subsubsection{{Numerics: methods to determine the critical point from $\tau$}} 
The critical point is determined from the dynamical response of the first-order coherence 
shown in Fig.~2\textbf{F} in the main text. For $|\varepsilon| \to 0$, $\tau(\varepsilon)$ displays a pronounced increase, before saturating due to finite size effects. 
We identify $\mathcal{P}_{\mathrm{th}}$ from the location of the maximum/incipient divergence of $\tau(\mathcal{P})$.
We then use the determined value of $\mathcal{P}_{\mathrm{th}}$ to define the reduced control parameter as $\varepsilon = \mathcal{P} / \mathcal{P}_{\mathrm{th}} - 1$.

To determine the center of the peak, we exploit the expected symmetry of the profile $\tau(\mathcal{P})$ around its maximum. We define the central point $\mathcal{P}_{\mathrm{th}}$ as the value that minimizes the asymmetry between the left and right sides of the peak. Specifically, $\mathcal{P}_{\mathrm{th}}$ is obtained by minimizing the functional
\begin{equation}
\mathcal{A}\!\left(\mathcal{P}_\mathrm{th}\right)
= \sum_{i}
\left[
\tau\!\left(\mathcal{P}_{\mathrm{th}} + \Delta_i\right)
-
\tau\!\left(\mathcal{P}_{\mathrm{th}} - \Delta_i\right)
\right]^2 ,
\end{equation}
where $\Delta_i$ spans the available range of normalized pump powers. This procedure allows us to extract the center of the peak without assuming a specific functional form for the underlying profile.

\

\subsubsection{Locating the critical point in trapped systems}

In numerical simulations, $\tau(P)$ can be reliably extracted on both sides of the transition and fitted to determine the dynamical critical point. Experimentally, the extraction of $\tau$ becomes increasingly difficult above threshold, making it useful to identify a complementary stationary observable that can locate the same transition. 

\begin{figure}
\centering
\includegraphics[width=\linewidth]{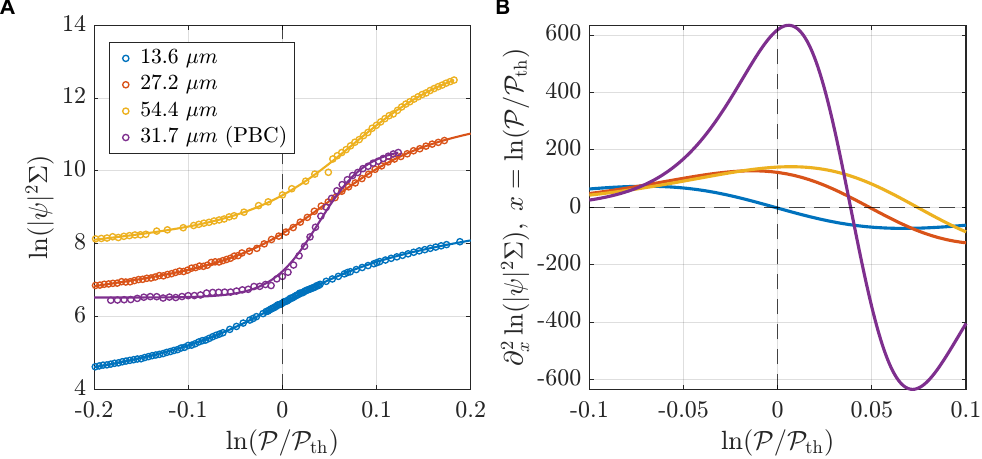
}
\caption{\textbf{Identification of the threshold in finite systems}.
\textbf{A} Integrated density $|\psi|^2 \Sigma$ as a function of normalized pump power $\mathcal{P}/\mathcal{P}_{\mathrm{th}}$ for different system sizes and boundary conditions. Here, $\mathcal{P}_\mathrm{th}$ is inferred from the divergence of the relaxation time $\tau$, as shown in Fig.~\ref{fig:finite_size_sims} (C). Solid lines show fits to Eq.~(\ref{eq:th_fit}), capturing the crossover across the transition.
\textbf{B} Curvature $d^2y/dx^2$ of the fitted density profiles (with $x=\ln(\mathcal{P}/\mathcal{P}_{\mathrm{th}})$ and $y=\ln(|\psi|^2 \Sigma)$), exhibiting a clear sign change at the threshold for finite trap sizes (here, the density-based criterion closely matches the dynamical threshold) while deviations emerge as the system size increases. 
}
\label{fig:th_finite_size_sys}
\end{figure}

We therefore use numerics to compare the critical point inferred from the dynamical fits with the behavior of the integrated polariton density.
Figure~\ref{fig:th_finite_size_sys}\textbf{A} shows the number of polaritons $|\psi|^2 \Sigma$ as a function of $\mathcal{P} / \mathcal{P}_\mathrm{th}$. 
Here, $\Sigma$ denotes the condensate area. For harmonically trapped configurations, we take $\Sigma = \pi d^2$, where $d$ is the characteristic radius of the condensate, whereas for periodic boundary conditions (PBC), we set $\Sigma = L^2$, with $L$ the linear size of the simulation domain.

We find that for the trap sizes considered in this work (e.g., $d \sim 14\,\mu\mathrm{m}$), the critical point closely coincides with a simple density-based criterion.
In particular, the pump value $\mathcal{P}_{\mathrm{th}}$ identified from the maximum of $\tau(\mathcal{P})$ coincides with the inflection point in log–log plots of the fitted integrated emission intensity as a function of pump power. 
This change of concavity is clearly visible in Fig.~\ref{fig:th_finite_size_sys}\textbf{B}, where the numerically-extracted curvature reverses sign at the critical point. 
However, this agreement is not universal. As the system size is increased, the dynamical threshold and the density-based criterion can separate, reflecting the evolution toward a sharper condensation transition. In the limit of large systems, both criteria tend toward the onset of nonlinear, superlinear density growth, which is commonly used to identify condensation in extended polariton systems.
Thus, while the correspondence is not guaranteed for arbitrary system size, the identification of the critical point at the change of concavity in the polariton density is quantitatively reliable in the finite-size regime explored here.

We now describe the numerical procedure used to fit the density growth across the transition (solid lines in Fig.~\ref{fig:th_finite_size_sys}\textbf{A}) and to compute the associated inflection point plotted in Fig.~\ref{fig:th_finite_size_sys}\textbf{B}.
We consider the steady-state density $|\psi|^2$ as a function of the pump $\mathcal{P}$ and work in logarithmic variables
\begin{equation}
x=\ln(\mathcal{P}/\mathcal{P}_{\rm th}),\qquad y=\ln(|\psi|^2 \Sigma).
\end{equation}  
To capture the crossover at threshold we fit $y(x)$ with a linear background plus a smooth transition,
\begin{equation}
y(x)=p_1+p_2 x+p_3\tanh\!\left(\frac{x-p_4}{p_5}\right),
\label{eq:th_fit}
\end{equation}
where $p_4$ and $p_5$ respectively set the center and width of the crossover.
To avoid bias from asymptotic regions, the fit is restricted to the interval where the slope change is maximal. 
To quantify the transition, we evaluate the curvature of the fitted model. Defining $u=(x-p_4)/p_5$, the second derivative reads
\begin{equation}
\frac{d^2y}{dx^2}
=-\frac{2p_3}{p_5^2}\tanh(u)\,\mathrm{sech}^2(u),
\label{eq:th_curv}
\end{equation}
which yields a characteristic sign change around $x\simeq p_4$. In this representation, $p_4$ provides an effective threshold shift in log-power, while $p_5$ and $p_3/p_5^2$ quantify the width and sharpness of the crossover.

\newpage

\end{document}